%% file: ANA-HMBS-2025-06-PAPER.tex
\pdfoutput=1
\documentclass[PAPER,coverpage=false, atlasdraft=false, UKenglish, cernpreprint, texmf, orcidlogo]{atlasdoc}
\usepackage[subfigure,abbrevref=false,multirow=true]{atlaspackage}
\usepackage{makecell}
\usepackage{comment}
\usepackage{atlasbiblatex}

\usepackage{atlasphysics}
\usepackage{atlasparticle}
\usepackage{atlasbsm}
\usepackage{pgffor}

\usepackage{xspace}
\usepackage{multirow}
\usepackage{textgreek}
\usepackage{adjustbox}
\usepackage{longtable}

\graphicspath{{logos/}{figures/}}

\usepackage{ANA-HMBS-2025-06-PAPER-defs}

\AtlasTitle{Search for direct pair production of top squarks in $pp$ collisions at $\sqrt{s}= 13$ TeV and $13.6$ TeV in events with two oppositely charged leptons using the ATLAS detector}

\AtlasAbstract{%
This paper presents the search for direct pair production of top squarks decaying into two on-shell top quarks and two neutralinos in final states with two oppositely charged leptons (electrons or muons), $b$-jets and large missing transverse momentum.
The search uses the full Run 2 dataset, corresponding to an integrated luminosity of 140~fb$^{-1}$ of proton--proton collisions at a centre-of-mass energy of $\sqrt{s} = 13$ TeV collected by the ATLAS detector from 2015 to 2018,
and the early Run 3 dataset, corresponding to an integrated luminosity of 53~fb$^{-1}$ of proton--proton collisions at $\sqrt{s} = 13.6$ TeV collected in 2022 and 2023.
Machine-learning-based classifiers are used to distinguish the signal from Standard Model backgrounds, to enhance signal discrimination across a wide kinematic phase space.
No significant excess is observed with respect to the Standard Model prediction and 95\% confidence level limits are set on top squark and neutralino mass combinations reaching up to 1060 GeV for the former and 560 GeV for the latter,
improving mass limits by about 10\% compared to previous ATLAS analyses in the same channel.

}

\AtlasRefCode{HMBS-2025-06}

\PreprintIdNumber{CERN-EP-2026-063}

\AtlasJournal{JHEP}
\AtlasCoverEgroupAnalysisTeam{atlas-hmbs-2025-06-analysis-team}

\hypersetup{pdftitle={ATLAS document},pdfauthor={The ATLAS Collaboration}}

\begin{document}

\maketitle

\section{Introduction}
%

Supersymmetry (SUSY)~\cite{Golfand:1971iw,Volkov:1973ix,Wess:1974tw,  Wess:1974jb, Ferrara:1974pu,Salam:1974ig,Farrar:1978xj} extends the Standard Model (SM)~\cite{DAmbrosio2002,Isidori2012} by predicting new bosonic partners for the existing fermions and fermionic partners for the known bosons. Of particular interest at the Large Hadron Collider (LHC) are searches for the supersymmetric partner of the top quark, hereafter referred to as the top squark.
The superpartners of the left- and right-handed top quarks, \stopL and \stopR, mix to form two mass eigenstates, \stopone and \stoptwo, where \stopone is the lighter state. The large top-quark Yukawa coupling can induce substantial mass splitting between these two eigenstates, implying that \stopone might be considerably lighter than the rest of the squarks.
Light top squarks are theoretically motivated because they cancel out the dominant quadratically divergent loop corrections from top quarks to the Higgs boson mass, making SUSY an appealing solution to the hierarchy problem~\cite{PhysRevD.13.974,PhysRevD.14.1667,PhysRevD.19.1277,PhysRevD.20.2619}. In a generic R-parity-conserving minimal supersymmetric extension of the SM (MSSM)~\cite{Farrar:1978xj,DIMOPOULOS1981150,Fayet:1976et,Fayet:1977yc,FAYET1979416}, SUSY particles are produced in pairs, and the lightest supersymmetric particle (LSP) is stable. The mass eigenstates formed from mixtures of the SUSY partners of the Higgs and electroweak gauge bosons are referred to as charginos (\chinoonepm, \chinotwopm) and neutralinos (\ninoone, \ninotwo, \ninothree, \ninofour), where the subscripts denote increasing mass.
The lightest neutralino (\ninoone) is here assumed to be the LSP and is a viable dark matter candidate because it is stable and interacts only weakly with ordinary matter~\cite{Goldberg:1983nd,Ellis:1983ew}.

\begin{figure}[h]
\centering
\includegraphics[width=0.3\linewidth]{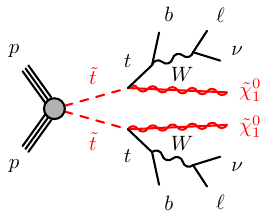}
\caption{Diagram of top-squark pair production, representing the two-body \stop decay into an on-shell top quark and the lightest neutralino ($\stop \rightarrow t \ninoone$).}
\label{fig:feyn_stop2b2L}
\end{figure}

This paper focuses on the search for direct pair production of \stopone particles,
that produce substantial missing transverse momentum in their decays due to undetected LSPs.
Scenarios in which the top squark decays into an on-shell top quark (which decays into \Wboson and \(b\)-quarks) and the lightest neutralino are considered, as depicted in~\cref{fig:feyn_stop2b2L}. Dedicated searches for direct \stopone pair production in the two-lepton final state were performed by the ATLAS~\cite{SUSY-2018-08} and CMS~\cite{CMS-SUS-19-011,} Collaborations using LHC \RunTwo data, with results excluding \stopone masses up to $1~\TeV$ at $95\%$ confidence level.
This analysis extends the search of Ref.~\cite{SUSY-2018-08} by including LHC \RunThr data collected in 2022 and 2023 and introducing a neural-network-based approach to enhance signal discrimination, but is restricted to on-shell top decays. For these decays, the dileptonic final state is particularly sensitive to models where the \ninoone is massive. Furthermore, it offers complementary sensitivity to analyses targeting the single lepton or the all-hadronic channels~\cite{SUSY-2018-12,SUSY-2018-07,SUSY-2019-12,SUSY-2023-22,CMS-SUS-19-010,CMS-SUS-19-009,CMS-SUS-20-002}.

%
%


%
\FloatBarrier
\section{ATLAS detector}
%

%
\newcommand{\AtlasCoordFootnote}{%
ATLAS uses a right-handed coordinate system with its origin at the nominal interaction point (IP) in the centre of the detector and the \(z\)-axis along the beam pipe.
The \(x\)-axis points from the IP to the centre of the LHC ring, and the \(y\)-axis points upwards.
Polar coordinates \((r,\phi)\) are used in the transverse plane, \(\phi\) being the azimuthal angle around the \(z\)-axis.
The pseudorapidity is defined in terms of the polar angle \(\theta\) as \(\eta = -\ln \tan(\theta/2)\) and is equal to the rapidity $ y = \frac{1}{2} \ln \left( \frac{E + p_z}{E - p_z} \right) $ in the relativistic limit.
Angular distance is measured in units of \(\Delta R \equiv \sqrt{(\Delta \eta)^{2} + (\Delta\phi)^{2}}\).}

The ATLAS detector~\cite{PERF-2007-01,GENR-2019-02} at the LHC covers nearly the entire solid angle around the collision point.\footnote{\AtlasCoordFootnote}
It consists of an inner tracking detector surrounded by a thin superconducting solenoid, electromagnetic and hadronic calorimeters, and a muon spectrometer incorporating three large superconducting air-core toroidal magnets.

The inner-detector system (ID) is immersed in a \qty{2}{\tesla} axial magnetic field
and provides charged-particle tracking in the range \(|\eta| < 2.5\).
The high-granularity silicon pixel detector covers the vertex region and typically provides four measurements per track,
the first hit generally being in the insertable B-layer.
It is followed by the SemiConductor Tracker, which usually provides eight measurements per track.
These silicon detectors are complemented by the transition radiation tracker (TRT),
which enables radially extended track reconstruction up to \(|\eta| = 2.0\).
The TRT also provides electron identification information
based on the fraction of hits (typically 30 in total) above a higher energy-deposit threshold corresponding to transition radiation.

The calorimeter system covers the pseudorapidity range \(|\eta| < 4.9\).
Within the region \(|\eta|< 3.2\), electromagnetic calorimetry is provided by barrel and
endcap high-granularity lead/liquid-argon (LAr) calorimeters,
with an additional thin LAr presampler covering \(|\eta| < 1.8\)
to correct for energy loss in material upstream of the calorimeters.
Hadronic calorimetry is provided by the steel/scintillator-tile calorimeter,
segmented into three barrel structures within \(|\eta| < 1.7\), and two copper/LAr hadronic endcap calorimeters.
The solid angle coverage is completed with forward copper/LAr and tungsten/LAr calorimeter modules
optimised for electromagnetic and hadronic energy measurements, respectively.

The muon spectrometer (MS) comprises separate trigger and
high-precision tracking chambers measuring the deflection of muons in a magnetic field generated by the superconducting air-core toroidal magnets.
The field integral of the toroids ranges between \num{2.0} and \qty{6.0}{\tesla\metre}
across most of the detector.
Three layers of precision chambers, each consisting of layers of monitored drift tubes, cover the region \(|\eta| < 2.7\).
These were complemented in the innermost layer of the endcap region by cathode-strip chambers in \RunTwo,
which were replaced in \RunThr by layers of small-strip thin-gap chambers and Micromegas chambers, both providing precision tracking in the region \(1.3 < |\eta| < 2.7\).
The muon trigger system covers the range \(|\eta| < 2.4\) with resistive-plate chambers in the barrel region, thin-gap chambers in the endcap regions,
and, in \RunThr, the aforementioned small-strip thin-gap chambers and Micromegas chambers in the innermost layer of the endcap.

The luminosity is measured mainly by the LUCID--2 detector that records Cherenkov light produced in the quartz windows of photomultipliers located close to the beam pipe.

Events were selected by the first-level trigger system implemented in custom hardware,
followed by selections made by algorithms implemented in software in the high-level trigger~\cite{TRIG-2016-01,TRIG-2022-01}.
The first-level trigger accepted events from the \qty{40}{\MHz} bunch crossings at a rate close to \qty{100}{\kHz},
which the high-level trigger further reduced in order to record complete events to disk at about \qty{1.25}{\kHz} and \qty{3}{\kHz} in \RunTwo and \RunThr, respectively.

The \RunThr detector configuration benefits from several upgrades compared with that of \RunTwo to maintain high detector performance at the higher \pileup levels of \RunThr.
The improvements include a new innermost layer of the muon spectrometer in the endcap region, which provides higher redundancy and a large reduction in fake muon triggers.
The trigger system also benefits from a new digital electronics readout of the LAr calorimeters with significantly increased granularity.
Other updates and further details are provided in Ref.~\cite{GENR-2019-02}.

A software suite~\cite{SOFT-2022-02} is used in data simulation, in the reconstruction and analysis of real and simulated data, in detector operations, and in the trigger and data acquisition systems of the experiment.


%
\FloatBarrier
\section{Data and simulated event samples}
%

%
The data used in this analysis were collected by the ATLAS detector during proton--proton (\(pp\)) collisions at centre-of-mass energies of $\sqrt{s} = 13~\TeV$ and $\sqrt{s} = 13.6~\TeV$ with a bunch-crossing separation of $25$~ns.
The $\sqrt{s} = 13~\TeV$ data sample corresponds to an integrated luminosity of $140~\ifb$, collected during the full \RunTwo of the LHC between 2015 and 2018. In this data-taking period, the number of interactions per bunch crossing ranged from about 8 to 70, with an average of 34~\cite{DAPR-2018-01}.
The $\sqrt{s} = 13.6~\TeV$ data sample was collected during the first two years (2022--2023) of \RunThr and corresponds to an integrated luminosity of $53~\ifb$.  During this period, the average number of interactions per bunch crossing increased to 42 and 51 for 2022 and 2023, respectively~\cite{ATL-DAPR-PUB-2024-001}.

Only events taken in stable beam conditions, and for which all relevant detector systems were operational, are considered~\cite{DAPR-2018-01}.
The quoted integrated luminosity is derived after the rejection of the data-taking periods that do not meet these detector requirements.
The uncertainties on the integrated luminosity are quoted to be $0.83\%$ and $2\%$ for the data collected in \RunTwo and \RunThr, respectively~\cite{DAPR-2021-01,ATL-DAPR-PUB-2024-001}.
The LUCID-2 detector~\cite{LUCID2} is used to obtain the primary bunch-by-bunch luminosity measurement, which is complemented by measurements obtained using the ID and calorimeters for the evaluation of the uncertainty.

Data events are required to satisfy either the single-electron or single-muon unprescaled trigger requirements~\cite{TRIG-2016-01,TRIG-2018-01,TRIG-2018-05,TRIG-2022-01}.
The triggering thresholds on the lepton transverse momentum (\pt) vary across data-taking periods, with a minimum threshold of $24$ ($26$) \GeV for electron triggers, and of $20$ ($24$) \GeV for muon triggers for \RunTwo (\RunThr).
To ensure a constant trigger efficiency, the leading lepton is required to have $\pT>\SI{27}{\GeV}$~\cite{TRIG-2018-05}.

The analysis makes use of Monte Carlo (MC) simulations to model the \SM backgrounds and the SUSY signals.
Additional MC samples are used to estimate systematic uncertainties, as detailed in Section~\ref{sec:systematics}.
The ATLAS simulation framework~\cite{SOFT-2010-01,SOFT-2022-02} was used to produce all MC samples, including a detailed simulation of the detector response based on \GEANT~\cite{Agostinelli:2002hh}.
The effect of multiple interactions in the same and neighbouring bunch crossings (\pileup) was modelled by overlaying~\cite{SIMU-2020-01} the simulated hard-scattering event with inelastic \(pp\) events generated from a mix of \EPOS[2.0.1.4]~\cite{Werner:2005jf} and \PYTHIA[8.308]~\cite{Bierlich:2022pfr}.
The \EPOS events were generated with the \EPOS LHC tune~\cite{Pierog:2013ria} and the \PYTHIA events with the A3 tune~\cite{ATL-PHYS-PUB-2016-017} and the \NNPDF[2.3lo]~\cite{Ball:2012cx, Ball:2014uwa} set of parton distribution functions (PDF).
\PYTHIA \pileup events include either a high \pt jet, a prompt photon, or a lepton from a $b$-hadron decay, while \EPOS was filtered to simulate all remaining \pileup events in the overlay sample.
The simulations obtained from the two generators were first reweighted to ensure a smooth transition across jet \pt. The combined sample was then reweighted to match the distribution of the actual number of interactions per bunch crossing measured in data.

The events targeted by this analysis are characterised by two charged leptons (electrons or muons), jets and missing transverse momentum.
The main SM background contributions are expected to come from top-quark pair production (\ttbar), associated production of a $Z$ boson and a top-quark pair pair (\ttZ), single-top in the $Wt$ production channel ($Wt$), \zjets production and diboson processes ($VV$ with $V=W,Z$). The set-up for the production of events and the cross-section computation accuracies used for the normalisation of these processes are summarised in Table~\ref{tab:samples_desc}.
Minor background processes such as $\ttbar t$, $\ttbar\ttbar$, $\ttbar W$, $\ttbar WW$, $\ttbar WZ$, $\ttbar H$, $tZ$ and $tWZ$ are simulated using the \MGNLO~\cite{Alwall:2014hca} generator interfaced with \PYTHIA~\cite{Sjostrand:2007gs} using the A14 tune~\cite{ATL-PHYS-PUB-2014-021} and the \NNPDF[2.3lo]~\cite{Ball:2014uwa} PDF set.
The SUSY top squark pair ($\stopone\stopone$) signal samples were generated from leading-order (LO) matrix elements with up to two extra partons using \MGNLO[3.5.3]~\cite{Alwall:2014hca}, interfaced to \MADSPIN~\cite{Sjostrand:2014zea,Artoisenet:2012st} and \PYTHIA[8.312]. %
Signal cross sections were calculated at next-to-next-to-leading order (NNLO) in the strong coupling constant, adding the resummation of soft gluon emission at next-to-next-to-leading-logarithm (NNLL) accuracy (giving \enquote{approximate NNLO+NNLL} accuracy)~\cite{Beenakker:2024nlf,Beenakker:2016lwe,Beenakker:2014sma,Beenakker:2013mva,Beenakker:2011sf,Beenakker:2009ha,Kulesza:2009kq,Kulesza:2008jb,Beenakker:1996ch}. %
The nominal cross section and its uncertainty were derived using the \PDFforLHC[21\_40\_pdfas] PDF set, following the recommendations of Ref.~\cite{Ball:2022oua}. Signal samples were generated for \stopone masses between 500 \GeV and 1300 \GeV and \ninoone masses up to 800 \GeV, both being varied in steps of $100~\GeV$.
\begin{table}[!h]
\centering
\caption{
Summary of the set-up for the production of events and the cross section computation accuracies used to simulate the signal and the main background processes.
The \enquote{dedicated} tune is the set of tuned parton-shower parameters developed by the \SHERPA authors. The symbol \enquote{$V$} stands for a vector boson.}
\label{tab:samples_desc}
\begin{adjustbox}{width=1\textwidth}
\begin{tabular}{llllllll}
\toprule
Physics process           & $\sqrt{s}$~[\TeV] & Generator                                                                        & Parton Shower                                            & Accuracy                                                                                                                                                                  & Tune                             & PDF (generator)                                      & PDF (shower)\\
\midrule
$Z(\ell\ell )$+jets       & $13$              & \SHERPA[2.2.11]~\cite{Bothmann:2019yzt}                                          & Dedicated                                                & NNLO~\cite{Anastasiou:2003ds}                                                                                                                                             & Dedicated                        & \NNPDF[3.0nnlo]~\cite{Ball:2014uwa}                  & \NNPDF[3.0nnlo]~\cite{Ball:2014uwa} \\
& $13.6$            & \SHERPA[2.2.14]~\cite{Bothmann:2019yzt}                                          & Dedicated                                                & NNLO                                                                                                                                                                      & Dedicated                        & \NNPDF[3.0nnlo]                                      & \NNPDF[3.0nnlo] \\
$Z(\tau\tau )$+jets       & $13/13.6$         & \SHERPA[2.2.14]                                                                  & Dedicated                                                & NNLO                                                                                                                                                                      & Dedicated                        & \NNPDF[3.0nnlo]                                      & \NNPDF[3.0nnlo] \\
$W(\ell\nu )$+jets        & $13/13.6$         & \SHERPA[2.2.11]/2.2.14                                                           & Dedicated                                                & NNLO                                                                                                                                                                      & Dedicated                        & \NNPDF[3.0nnlo]                                      & \NNPDF[3.0nnlo] \\
$VV$, $VVV$               & $13$              & \SHERPA[2.2.1],\,2.2.2,\,2.2.14                          & Dedicated                                                & NLO~\cite{Hoeche:2011fd,Hoeche:2012yf,Catani:2001cc,Hoeche:2009rj}                                                                                                        & Dedicated                        & \NNPDF[3.0nnlo]                                      & \NNPDF[3.0nnlo] \\
& $13.6$            & \SHERPA[2.2.14]                                                                  & Dedicated                                                & NLO                                                                                                                                                                       & Dedicated                        & \NNPDF[3.0nnlo]                                      & \NNPDF[3.0nnlo] \\
$t\bar{t}$                & $13/13.6$         & \POWHEGBOX[v2]~\cite{Frixione:2007nw,Nason:2004rx,Frixione:2007vw,Alioli:2010xd} & \PYTHIA[8]~\cite{Sjostrand:2014zea}                      & NNLO+NNLL~\cite{Beneke:2011mq,Cacciari:2011hy,Baernreuther:2012ws,Czakon:2012zr,Czakon:2012pz,Czakon:2013goa,Czakon:2011xx}                                               & A14~\cite{ATL-PHYS-PUB-2014-021} & \NNPDF[3.0nlo]~\cite{Ball:2014uwa}                   & \NNPDF[2.3lo]~\cite{Ball:2012cx, Ball:2014uwa}\\
Single-top: \\
$~~~~Wt$                  & $13/13.6$         & \POWHEGBOX[v2]                                                                   & \PYTHIA[8]                                               & NLO+NNLL~\cite{Kidonakis:2010ux,Kidonakis:2013zqa}                                                                                                                        & A14                              & \NNPDF[3.0nlo]                                       & \NNPDF[2.3lo]    \\
~~~~$s$- and $t$-channels & $13/13.6$         & \POWHEGBOX[v2]                                                                   & \PYTHIA[8]                                               & NLO~\cite{Aliev:2010zk,Kant:2014oha}                                                                                                                                      & A14                              & \NNPDF[3.0nlo]                                       & \NNPDF[2.3lo]      \\
$\ttZ$                    & $13/13.6$         & \MGNLO[2.3.3]~\cite{Alwall:2014hca}                                              & \PYTHIA[8]                                               & NLO~\cite{deFlorian:2016spz}                                                                                                                                              & A14                              & \NNPDF[3.0nlo]                                       & \NNPDF[2.3lo] \\
$\ttW$                    & $13/13.6$         & \SHERPA[2.2.10]/2.2.14                                  & Dedicated                                                & NLO                                                                                                                                                                      & Dedicated                              & \NNPDF[3.0nnlo]                                      & \NNPDF[3.0nnlo]     \\
\midrule
SUSY signal               & $13/13.6$         & \MADGRAPH[3.5.3]~\cite{Alwall:2014hca}                                           & \PYTHIA[8]                                               & NNLO+NNLL~\cite{Beenakker:2024nlf,Beenakker:2016lwe,Beenakker:2014sma,Beenakker:2013mva,Beenakker:2011sf,Beenakker:2009ha,Kulesza:2009kq,Kulesza:2008jb,Beenakker:1996ch} & A14                              & \NNPDF[3.0nlo]                                       & \NNPDF[2.3lo]  \\
\bottomrule
\end{tabular}
\end{adjustbox}
\end{table}


%
\FloatBarrier
\section{Event reconstruction}
\label{sec:object-identification}

%
%

Primary vertices are reconstructed from ID tracks with \(\pt>\SI{500}{\MeV}\)~\cite{ATL-PHYS-PUB-2015-026,ATL-PHYS-PUB-2019-015}.
The primary vertex with the largest sum of squared track \pt is chosen as the hard-scatter vertex, and serves as the origin for all the objects described below.

Leptons (electrons or muons) are classified as either baseline or signal.
Baseline leptons are used in the calculation of the missing transverse momentum, to resolve ambiguities between the analysis objects in an event, and to define 2-lepton and 3-lepton analysis regions.
Baseline electrons are reconstructed from energy deposits (clusters) in the electromagnetic calorimeter matched to tracks in the ID~\cite{PERF-2017-01,EGAM-2018-01,EGAM-2021-02}.
Baseline electron candidates are required to have a pseudorapidity $|\eta|<2.47$, a transverse energy $\ET>\SI{4.5}{\GeV}$, to satisfy a \textit{Loose} likelihood-based identification requirement~\cite{EGAM-2021-01}, and to have $|z_0 \sin\theta| < 0.5$~mm, where \(z_0\) is the longitudinal impact parameter relative to the reconstructed primary vertex.
They are calibrated based on the procedure and results described in Ref.~\cite{EGAM-2018-01,EGAM-2021-02}. Additional corrections to the in-situ energy scale and resolution are applied to account for residual differences between data and simulation.
Baseline muon candidates are reconstructed by matching ID tracks in the pseudorapidity region $|\eta|<2.4$ with MS tracks and energy deposits in the calorimeter compatible with a minimum-ionising particle.
The resulting tracks are  required to have a  $\pT>\SI{4}{\GeV}$ and to satisfy $|z_0 \sin\theta| < 0.5$~mm.
Baseline muon candidates are required to satisfy the \textit{Medium} identification requirement, defined in Ref.~\cite{PERF-2015-10}, based on the number of hits in the different ID and MS subsystems, and on the precision of the charge-to-momentum ratio $q/p$ measurement.

Jets are reconstructed using the anti-$k_t$ algorithm~\cite{Cacciari:2008gp} as implemented in FastJet~\cite{Fastjet} with a jet radius parameter of $R=0.4$. The inputs to this algorithm are particle flow (PFlow) objects~\cite{PERF-2015-09}, which combine measurements from the ATLAS ID and calorimeters~\cite{PERF-2014-07} to improve the jet energy resolution and increase the jet reconstruction efficiency, especially at low jet \pT.
The reconstructed jets are then calibrated by the application of a jet energy scale and resolution correction derived from data and simulation~\cite{JETM-2018-05}.
Only jet candidates with $\pT>\SI{25}{\GeV}$ and $|\eta|<2.5$ are considered.\footnote{No attempt is made to identify hadronic $\tau$-lepton decays, so they are treated as jets.} To reduce the effects of \pileup, jets with $\pT < \SI{120}{\GeV}$  are required to satisfy the jet-to-vertex tagger~(JVT)~\cite{PERF-2014-03} \enquote{Tight} working point requirements. The tagger is configured to identify jets from the hard-scatter vertex with 92\% efficiency.
Jets originating from the hadronisation of $b$-quarks are identified using the GN2 $b$-tagging algorithm~\cite{FTAG-2019-07,FTAG-2023-05} at a working point corresponding to a $77\%$ efficiency for \bjets, with a misidentification rate below $0.06$ ($0.002$) for charm (light-flavour) jets in simulated \ttbar\ events. The \(b\)-tagging efficiencies in simulated events are corrected to match those measured in data~\cite{FTAG-2018-01,FTAG-2020-08, FTAG-2019-02}. Jets failing this \(b\)-tagging selection are hereafter called light-jets.

To avoid reconstruction ambiguities and double counting of analysis objects, an overlap removal procedure is applied to the baseline leptons and jets in the following order.
First, calorimeter-tagged muons, reconstructed from a track in the ID matched to an energy deposit in the calorimeter compatible with a minimum-ionising particle, are removed if sharing the track with electrons and, next, all electrons sharing an ID track with a muon are removed.
Jets that lie within a cone of $\Delta R = 0.2$ around an electron candidate are removed.
Finally,  any  lepton candidate is removed in favour of a jet candidate if it is closer than $\Delta R < \mathrm{min }(0.4, 0.04 + 10~\GeV/{p_{\mathrm T}^\ell})$ to a jet, where  ${p_{\mathrm T}^\ell}$ is the \pT of the lepton.

The missing transverse momentum ($\mathbf p^\mathrm{miss}_\mathrm{T}$), with magnitude \MET, is defined as the negative vector sum of the transverse momenta of all  baseline electrons, muons and jets.
Low-momentum tracks from the primary vertex that are not associated with reconstructed analysis objects are also considered in the calculation, by including them in a dedicated soft term~\cite{PERF-2016-07}.
Linked to the \MET is the `object-based \MET significance'~\cite{ATLAS-CONF-2018-038}, referred to as `\METsig' in this paper.
This quantity measures the  significance of the \MET based on the transverse momentum resolution of all objects used in the calculation of the $\mathbf p^\mathrm{miss}_\mathrm{T}$.
The \METsig improves the discrimination between events with true \MET, arising from weakly interacting particles such as neutralinos, and those where the \MET is consistent with a particle mismeasurement and resolution or identification inefficiencies, thus providing better background rejection. %

Signal leptons are a subset of the baseline leptons, satisfying stricter reconstruction quality and kinematic selections, and are used for the final event selection.
Signal electrons are required to satisfy a \textit{Tight} likelihood-based identification requirement~\cite{EGAM-2021-01} and the track associated with a signal electron is required to have a significance $\vert d_0\vert/\sigma(d_0) < 5$, where $d_0$ is the transverse impact parameter relative to the reconstructed primary vertex and $\sigma(d_0)$ is its uncertainty.
To further reduce the background from non-prompt leptons and mis-identified hadrons, signal electrons are also required to be isolated in the tracking system and the calorimeter, as described in Ref.~\cite{EGAM-2018-01}.
For signal muons, a significance in the transverse impact parameter  $\vert d_0\vert/\sigma(d_0) < 3$ is required.
Isolation criteria applied to muons require low activity in the tracker around the muon track, as well as in the calorimeters~\cite{MUON-2018-03}.


%
\FloatBarrier
\section{Event selection and analysis strategy}%
\label{sec:analysis-strategy}

The search reported in this paper selects events with a pair of leptons with opposite electric charge.
Events are required to contain at least two jets, with one identified as a \bjet.
The leading lepton is required to have a $\pt>27\,\GeV$ to be in the plateau region of the used triggers, while the subleading lepton is required to have a $\pt>20\,\GeV$.
A selection is also applied to the invariant mass of the two leptons, $\mll>20\,\GeV$, to reject events coming from low-mass resonances.

Six Neural Networks (NNs) are trained on MC simulated samples of the main SM backgrounds (\ttbar and \ttZ) and the signal models, only including events satisfying the selection described above.
The signal models are grouped into three categories according to the mass splitting $\DeltaM$:
\begin{itemize}
\item{\bfseries low-\DeltaM\ category}: comprising signal models with $\DeltaM =200-300 ~\GeV$;
\item{\bfseries medium-\DeltaM\ category}: comprising signal models with $\DeltaM =300-500 ~\GeV$;
\item{\bfseries high-\DeltaM\ category}: comprising signal models with $\DeltaM \geq  600 ~\GeV$.
\end{itemize}
This categorisation accounts for differences in the signal kinematics that are driven by \DeltaM, thereby providing wide-ranging discrimination between the signal and the backgrounds across the parameter space.
Two NNs are trained for each \DeltaM\ category: one using events with two or more $b$-tagged jets, and one using events with exactly one $b$-tagged jet,
to account for cases where the \ttbar system cannot be fully reconstructed.
An additional preselection on the stransverse mass \mttwoL~\cite{Lester_1999,Barr_2003} is applied prior to training to restrict the training data-set to the kinematic phase-space of interest for the targeted signal models, and depends on the \DeltaM\ category: $\mttwoL>80~\GeV$ for the low-\DeltaM\ category, $\mttwoL>90~\GeV$ for the medium-\DeltaM\ category, and $\mttwoL>100~\GeV$ for the high-\DeltaM\ category.
The variable \mttwo\ is  used to bound the masses of a pair of identical particles that have each decayed into a visible and an invisible particle.
Different variants of \mttwo can be computed, using different combinations of visible objects. For the \mttwo calculation, the invisible particles are assumed to be massless. The two signal leptons are used for the calculation of \mttwoL.
In dileptonic events, the \mttwoL\ distribution is expected to have an endpoint corresponding to the $W$ boson mass for backgrounds such as \ttbar, where neutrinos are the only source of \MET, while it is expected to reach higher values in the case of SUSY events, due to the presence of the neutralinos~\cite{Cho:2007dh,Burns:2008va}.

The NN input variables include the kinematic variables (\pT, $\eta$, $\phi$) of the signal leptons and the two most energetic jets in the final state, complemented by a set of additional observables that provide further discrimination between the background and the signal, which are defined in the following.
The lepton variables include the lepton flavour and the invariant mass (\mll) of the two leptons. For jets, the pseudo-continuous $b$-tagging score is also considered: it consists of the discrete output of the GN2 $b$-tagging algorithm (see \cref{sec:object-identification}), so that each bin corresponds to a given $b$-tagging efficiency range ($65\%$, $70\%$, $77\%$, $85\%$, and $90\%$).\\
At event-level, the input variables include the \met, with its significance and its azimuthal direction.
The angular separation ($\Delta\phi$) between all the combinations of two considered final state objects (the two leptons, the leading and the subleading \bjet) are considered, together with the transverse mass ($m_{\mathrm{T}}$) between the missing transverse momentum and each of the two leptons.
Additionally, the variable \pbll is defined as the vectorial sum of $\mathbf p^{\mathrm{miss}}_{\mathrm{T}}$ and the leptons transverse momentum vectors $\mathbf \pT(\ell_1)$  and  $\mathbf \pT(\ell_2)$.
Its magnitude, \ptllboost, can be interpreted as the magnitude of the vector sum of all the transverse hadronic activity in the event.
The azimuthal angle between the $ \mathbf p^{\mathrm{miss}}_{\mathrm{T}} $ vector and the \pbll\ vector is defined as \dphib.
This variable is useful for selecting events where the two signal leptons are in the same direction of the $\mathbf p^\mathrm{miss}_\mathrm{T}$, and is used together with \ptllboost as an input to the NNs~\cite{Polesello:2009rn}.

Finally, different variants of \mttwo\ are considered, including all the possible combinations of leptons and $b$-tagged jets as visible particles in the calculation of the variable.
For the calculation of the variables described above, when two $b$-tagged jets are required and only one is reconstructed, the leading light-jet is used instead of the subleading $b$-tagged jet.
According to metrics provided by the SHAP package~\cite{shap-values}, the input variables with best signal separation power used in the training were found to be \mttwoL\ and \met\ significance.
The NNs are trained using Keras~\cite{chollet2015keras} with the TensorFlow~\cite{abadi2016tensorflow} backend on a combined dataset of \RunTwo and \RunThr simulated samples.
The implemented models are multiclass classification NNs, with one class corresponding to the signal and classes for each of the dominant background processes (\ttbar\ and \ttZ).
The outputs of the NNs are classification scores ranging between $0$ and $1$ for each class, representing the probability that an event belongs to the given class. The NN signal and \ttbar\ scores, corresponding to the classifier outputs for the signal and \ttbar\ classes, respectively, are used in the definition of the analysis regions.

The signal regions (SRs) are then defined by applying a selection on the NN signal score, in addition to the preselection defined above, as summarised in \cref{tab:SR_selection}.
\begin{table}[!ht]
\centering
\caption{Definition of the signal regions used for the model-dependent interpretation. The variables \nbjet and \nljet represent the number of $b$-tagged jets and the number of light-flavour jets in an event, respectively.}
\label{tab:SR_selection}
\renewcommand{\arraystretch}{1.2}{\footnotesize
\begin{tabular}{l c c   c c  c c}
\toprule
& \multicolumn{2}{c}{\textbf{Low-$\Delta m$}} & \multicolumn{2}{c}{\textbf{Medium-$\Delta m$}} & \multicolumn{2}{c}{\textbf{High-$\Delta m$}}  \\
& \textbf{1b} & \textbf{2b} & \textbf{1b} & \textbf{2b} & \textbf{1b} & \textbf{2b} \\
\midrule
\ptleplead[\GeV] & \multicolumn{6}{c}{$>27$}\\
\ptlepsubl[\GeV] & \multicolumn{6}{c}{$>20$}\\
\mll[\GeV] & \multicolumn{6}{c}{$>20$}\\
\midrule
\nbjet & $=1$ & $>1$ & $=1$ & $>1$ & $=1$ & $>1$ \\
\nljet & $>0$ & $\ge0$ & $>0$ & $\ge0$ & $>0$ & $\ge0$ \\
\midrule
\mttwoL[\GeV]   & \multicolumn{2}{c}{$>80$} & \multicolumn{2}{c}{$>90$} & \multicolumn{2}{c}{$>100$} \\
\midrule
NN signal score & \multicolumn{2}{c}{$>0.93$} & \multicolumn{2}{c}{$>0.53$} & \multicolumn{2}{c}{$>0.44$} \\
\bottomrule
\end{tabular}
}
\end{table}
\cref{fig:NNscores} shows the expected event yield as a function of the NN signal score for the \RunThr MC samples in the three \DeltaM\ regions for the SM background processes and for three representative signal models: $m(\stopone,\ninoone) = (800, 500)~\GeV$ for the low-\DeltaM\ category, $m(\stopone,\ninoone) = (900, 500)~\GeV$ for the medium, and $m(\stopone,\ninoone) = (1000, 1)~\GeV$ for the high. %
The binning along the NN output variable is optimised to maximise the sensitivity across the (\stopone,\ninoone) mass plane individually for each NN to account for the different score distributions.
\begin{figure}[htb!]
\begin{center}
\subfigure[]{
\includegraphics[width=0.31\textwidth]{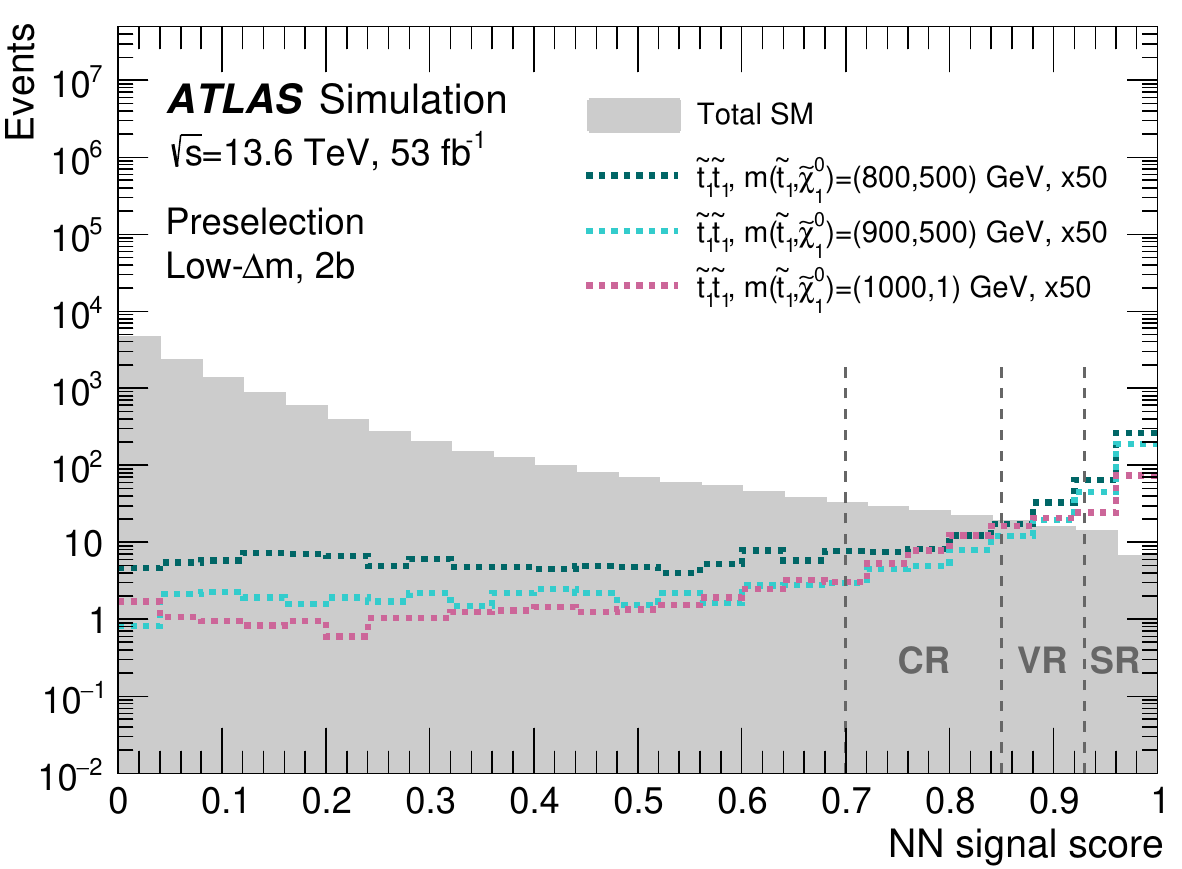}
}
\subfigure[]{
\includegraphics[width=0.31\textwidth]{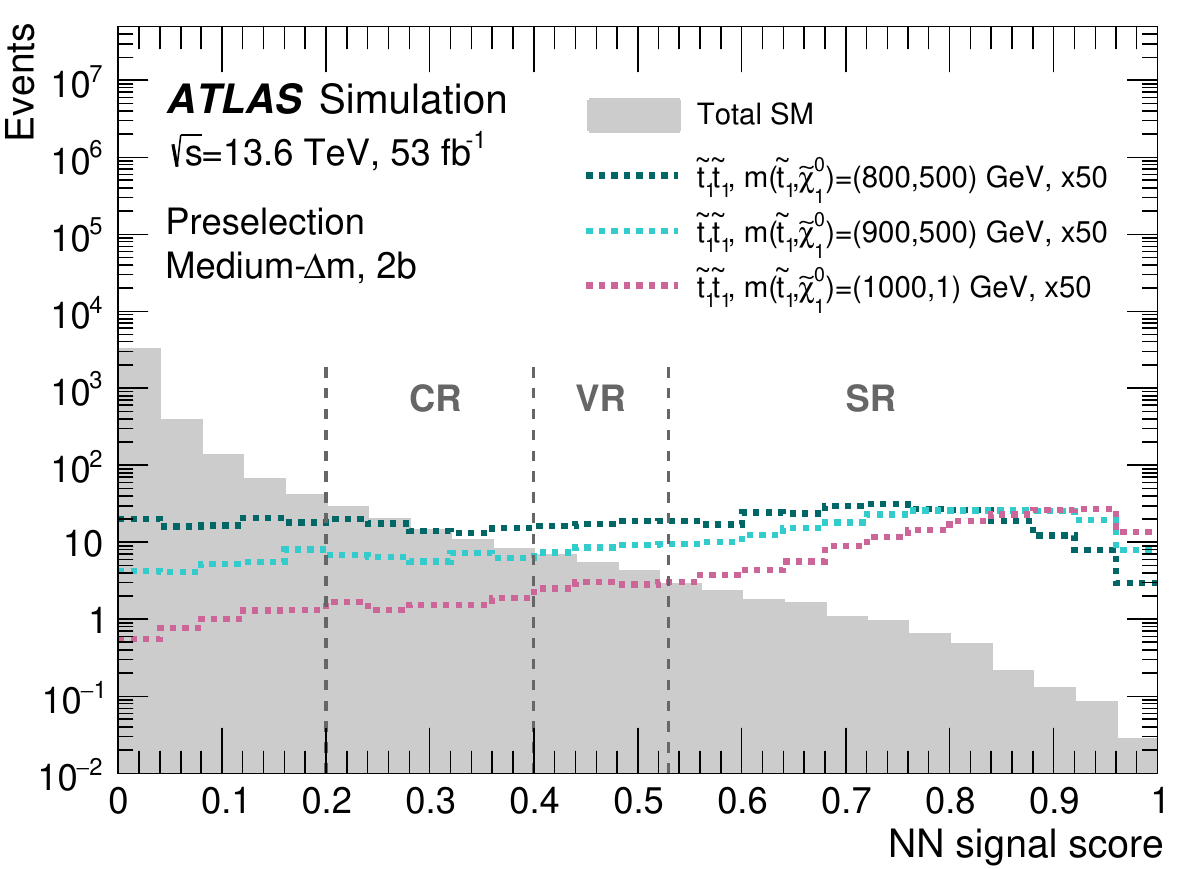}
}
\subfigure[]{
\includegraphics[width=0.31\textwidth]{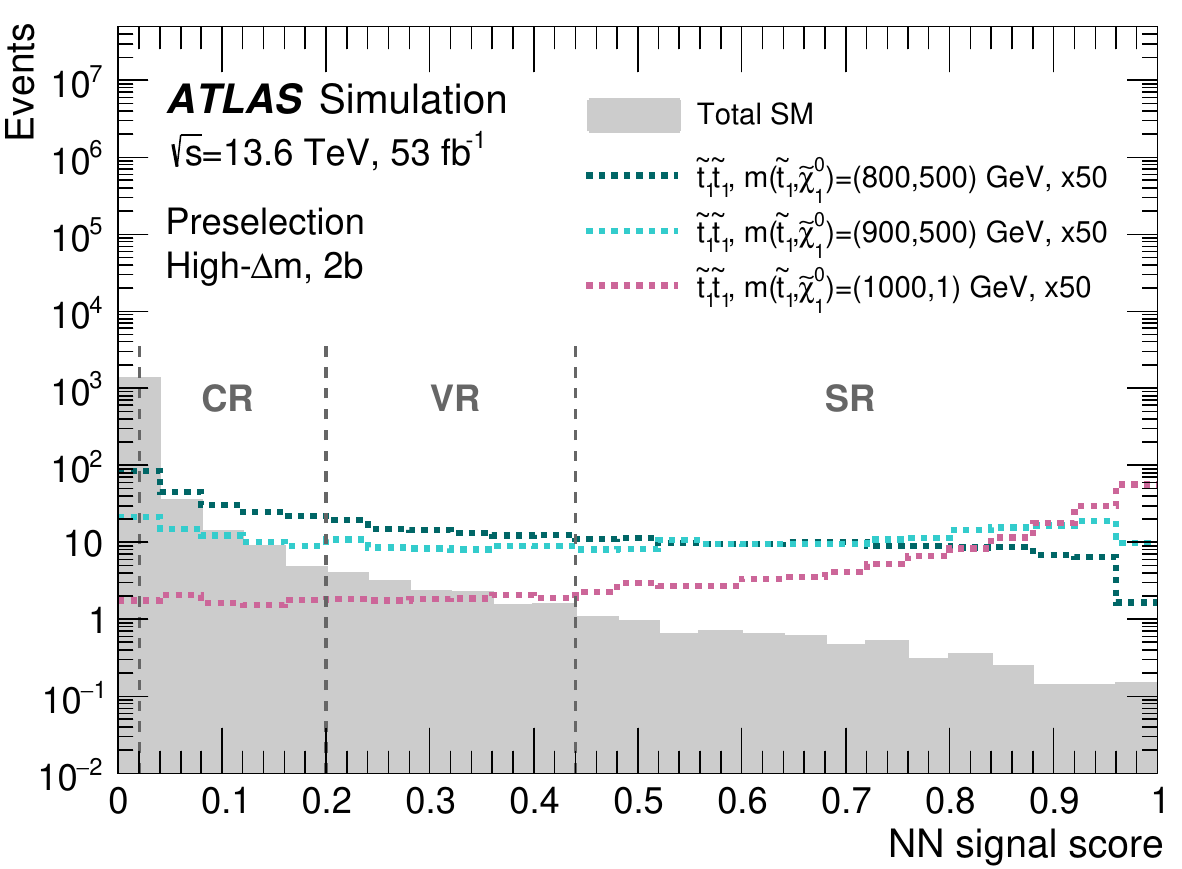}
}
\end{center}
\caption{Expected event yield as a function of the NN signal score for the (a) low, (b) medium, and (c) high \DeltaM\ regions. The scores are shown for \RunThr simulated events with two $b$-tagged jets.
The distribution in grey represents the expected number of SM background events, the dashed lines show the distributions of three representative signal models: $m(\stopone,\ninoone) = (800, 500)~\GeV$ for the low-\DeltaM\ category, $m(\stopone,\ninoone) = (900, 500)~\GeV$ for the medium, and $m(\stopone,\ninoone) = (1000, 1)~\GeV$ for the high. The expected signal yields are multiplied by 50.
CRs (Control Regions), and VRs (Validation Regions) are defined in Section~\ref{sec:background}.}
\label{fig:NNscores}
\end{figure}


%
\FloatBarrier
\section{Background estimation}
\label{sec:background}

The MC predictions for the dominant SM background processes \ttbar\ and \ttZ, with an invisible decay of the $Z$ boson ($Z\rightarrow \nu\bar{\nu}$), are improved using a data-driven normalisation procedure, while non-dominant processes are estimated directly using  MC simulation.
MC simulation is also used to account for background events arising from processes that do not contain two leptons at scattering level, but for which two leptons are selected by the analysis criteria, either because of hadrons
misidentified as leptons or non-prompt leptons from decays of hadrons (mainly from semileptonic \ttbar events).
The modelling of the processes with one or more fake or non-prompt leptons (hereafter referred to as FNP) was validated in data using a sample with two same-charge signal leptons, which is naturally enriched in this background.

For each \DeltaM\ category, a simultaneous profile likelihood fit~\cite{Rolke:2004mj} is used to constrain the MC yields with the observed data in dedicated background control regions (CRs).
The CRs are kinematic regions enriched in the targeted background. Their definitions aim at keeping the contamination of other backgrounds and signals as small  as possible, while being kinematically close to the SRs to reduce the extrapolation error.
The CRs and the SRs are required to be orthogonal to each other. %
In the fit, implemented with the HistFitter and PyHF frameworks~\cite{Baak:2014wma, pyhf_joss, pyhf_v0.7.2_zenodo}, the normalisation of the targeted backgrounds are allowed to float in an independent way for the \RunTwo and \RunThr datasets, while the MC simulation is used to describe the distribution of kinematic variables. %
Systematic uncertainties that could affect the expected yields in the different regions are taken into account in the fit through nuisance parameters.
Each uncertainty source is described by a single nuisance parameter constrained by a Gaussian distribution whose width corresponds to the size of the uncertainty, which is shared across analysis regions and background processes, when the related uncertainties are considered correlated.
A list of the systematic uncertainties considered in the fit is provided in Section~\ref{sec:systematics}.
The SM background estimate is validated in dedicated validation regions (VRs), which are disjoint from both CRs and SRs.

The \ttbar\ CRs and VRs are defined applying selections to the NN scores for each \DeltaM\ category. Consequently, different CRs and VRs are defined for events with one and more than one $b$-tagged jet.
Specifically, CRs contain events with lower NN signal score, and the VRs cover the gap between the CRs and the SRs, as depicted in~\cref{fig:NNscores}.
A selection on the NN \ttbar\ score is added to the definition of the \ttbar\ CRs, to remove the events with NN \ttbar\ score close to zero, thus increasing the purity of the region for the \ttbar\ process. %

\ttZ\ events with invisible decays of the $Z$ boson are expected to dominate the tail of the \mttwoL\ distribution in the SRs and are normalised in the dedicated control region \CRttZ.
Given the difficulty in achieving enough purity for this SM process because of the high contamination from \ttbar events, a strategy based on a three-lepton final state is adopted.
Events are selected if characterised by three charged leptons including at least one pair of same-flavour leptons with opposite-sign electric charge (SFOS leptons) having an invariant mass consistent with that of the $Z$~boson ($|\mll-m_{Z}|<20$~\GeV). If more than one pair is identified, the one with $\mll$ closest to the $Z$~boson mass is chosen. Events are further required to have a jet multiplicity, \njet, greater than or equal to three with at least two $b$-tagged jets.
These selections target \ttZ\ production with the $Z$ boson decaying into two leptons and \ttbar\ decaying in the semileptonic channel. %
To select \ttZ\ events whose kinematics, regardless of subsequent \ttbar\ and $Z$ decays, emulate the kinematics of this background in the SRs, the momenta of the two leptons of the SFOS pair $(\mathbf p(\ell^\mathrm{Z}_1),\mathbf p(\ell^\mathrm{Z}_2))$ are vectorially added to the $\mathbf p^\mathrm{miss}_\mathrm{T}$, effectively treating them as a neutrino pair from the $Z$ boson decay. A variable  called $\METcorr=\left|\left(\mathbf p^\mathrm{miss}_\mathrm{T}+\mathbf p(\ell^\mathrm{Z}_1)+\mathbf p(\ell^\mathrm{Z}_2)\right)_{\mathrm{T}}\right|$ is constructed.
Events characterised by high \MET selected by the NN scores in the SRs are emulated by requiring $\METcorr\  >  140~\GeV$ in \CRttZ. %

The expected signal contamination in the CRs is generally below $\sim 1\%$ and less than $10\%$ in the VRs. %
\cref{tab:CRyieldsRun2,tab:CRyieldsRun3} present the post-fit expected and observed yields in the previously described CRs, together with the normalisation factors (\(\mu\)) extracted for each \DeltaM\ category, for \RunTwo and \RunThr respectively.
The normalisation factor \(\mu_{\ttbar}\) assumes different values in the two datasets, because the \RunThr event selection is on average selecting events in which the \ttbar system is more energetic, due to the higher centre-of-mass energy of the collisions.
Only one representative version of \(\mu_{\ttZ}\) is listed for each dataset, as a common CR is defined for this background, and all the fits give similar results for this parameter.
\cref{fig:CRs} shows the agreement between data and background predictions for the post-fit distributions of the \MET and the \METcorr in \CRttbar{Medium,2b}\ and in \CRttZ\ using \RunThr data.

Finally, \cref{fig:VRpull} shows the post-fit event yields in the \ttbar\ VRs defined for the \RunTwo and \RunThr data-taking periods, respectively. Good agreement is generally found between the data and the SM expectation.
The only notable exception is the \RunThr medium-\DeltaM\ region with two $b$-tagged jets (\VRttbar{Medium,2b}), where an excess with a significance of more than two standard deviations is observed. A study of the events selected revealed no anomalous features.

\begin{table}[hptb]
\centering
\caption{Post-fit yields in the \RunTwo CRs. All systematic uncertainties as described in Section~\ref{sec:systematics} are included.
``Others'' includes  the contributions from $VVV$, $\ttbar t$, $\ttbar\ttbar$, $\ttbar W$, $\ttbar WW$, $\ttbar WZ$, $\ttbar H$, $tZ$ and $tWZ$ processes.}
\label{tab:CRyieldsRun2}
\begin{adjustbox}{max height=\textheight, max width=\textwidth}
\footnotesize
\renewcommand{\arraystretch}{1.1}
\begin{tabular}{lccccccc}
\toprule
& \CRttbar{Low,1b}             & \CRttbar{Low,2b}             & \CRttbar{Medium,1b}          & \CRttbar{Medium,2b}          & \CRttbar{High,1b}        & \CRttbar{High,2b}        & \CRttZ                      \\
\midrule
Observed                        & $232$                         & $193$                         & $219$                         & $151$                         & $345$                     & $207$                     & $412$                 \\
Total SM                        & $232 \pm 15$                  & $193 \pm 13$                  & $219 \pm 14$                  & $151 \pm 11$                  & $345 \pm 18$              & $206 \pm 13$              & $412 \pm 20$          \\
\midrule
\ttbar                          & $201 \pm 14$                  & $174 \pm 14$                  & $182 \pm 12$                  & $138 \pm 12$                  & $266 \pm 16$              & $176 \pm 14$              & $-$       \\
$Wt$                            & $20 \pm 4$                    & $11 \pm 1$                    & $23 \pm 3$                    & $7.31 \pm 0.64$               & $36 \pm 3$                & $15 \pm 1$                & $-$       \\
$Z/\gamma$ + jets               & $0.74 \pm 0.64$               & $0.71 \pm 0.23$               & $4.7 \pm 1.1$                 & $0.61 \pm 0.19$               & $19 \pm 2$                & $3.51 \pm 0.48$           & $-$       \\
Diboson                         & $1.46 \pm 0.44$               & $0.08_{-0.08}^{+0.09}$        & $1.64 \pm 0.35$               & $0.00_{-0.00}^{+0.04}$        & $5.09 \pm 0.56$           & $0.25 \pm 0.06$           & $52 \pm 7$      \\
\ttZ                            & $3.41 \pm 0.50$               & $3.08 \pm 0.46$               & $2.43 \pm 0.43$               & $1.67 \pm 0.34$               & $6.5 \pm 1.2$           & $3.88 \pm 0.59$           & $274 \pm 27$          \\
Others                          & $3.50 \pm 0.34$               & $2.90 \pm 0.25$               & $3.77 \pm 0.31$               & $2.55 \pm 0.30$               & $8.30 \pm 0.66$           & $5.52 \pm 0.28$           & $67 \pm 7$        \\
FNP                             & $1.95 \pm 0.67$               & $1.22 \pm 0.43$               & $1.86 \pm 0.42$               & $0.98 \pm 0.29$               & $3.90 \pm 0.75$           & $1.99 \pm 0.31$           & $19 \pm 2$        \\
\midrule
$\mu$                              & \multicolumn{2}{c}{$\mu_{\ttbar} = 0.86 \pm 0.05$}                              & \multicolumn{2}{c}{$\mu_{\ttbar} = 0.91 \pm 0.06$}                              & \multicolumn{2}{c}{$\mu_{\ttbar} = 0.95 \pm 0.05$}          & $\mu_{\ttZ} = 1.08 \pm 0.10$       \\
\bottomrule
\end{tabular}
\end{adjustbox}
\end{table}
\begin{table}[hptb]
\centering
\caption{Post-fit yields in the \RunThr CRs. All systematic uncertainties as described in Section~\ref{sec:systematics} are included.
``Others'' includes  the contributions from $VVV$, $\ttbar t$, $\ttbar\ttbar$, $\ttbar W$, $\ttbar WW$, $\ttbar WZ$, $\ttbar H$, $tZ$ and $tWZ$ processes.}
\label{tab:CRyieldsRun3}
\begin{adjustbox}{max height=\textheight, max width=\textwidth}
\footnotesize
\renewcommand{\arraystretch}{1.1}
\begin{tabular}{lccccccc}
\toprule
& \CRttbar{Low,1b}              & \CRttbar{Low,2b}              & \CRttbar{Medium,1b}           & \CRttbar{Medium,2b}           & \CRttbar{High,1b}         & \CRttbar{High,2b}         & \CRttZ                \\
\midrule
Observed                        & $113$                         & $106$                         & $107$                         & $70$                          & $174$                     & $116$                     & $163$                 \\
\midrule
Total SM                        & $115 \pm 10$                  & $104 \pm 9$                   & $107 \pm 9$                   & $70 \pm 7$                    & $177 \pm 12$              & $113 \pm 9$               & $163 \pm 13$          \\
\midrule
\ttbar                          & $103 \pm 10$                  & $97 \pm 9$                    & $91 \pm 9$                    & $65 \pm 7$                    & $145 \pm 12$              & $101 \pm 10$              & $-$       \\
$Wt$                            & $7.9 \pm 1.7$                 & $4.51 \pm 0.64$               & $9.9 \pm 1.8$                 & $2.93 \pm 0.43$               & $14 \pm 3$                & $6.51 \pm 0.90$           & $-$       \\
$Z/\gamma$ + jets               & $0.28_{-0.28}^{+0.32}$        & $0.04_{-0.04}^{+0.08}$        & $2.39 \pm 0.73$               & $0.34 \pm 0.14$               & $8.2 \pm 2.5$             & $1.33 \pm 0.34$           & $-$       \\
Diboson                         & $0.61 \pm 0.11$               & $0.08 \pm 0.02$               & $0.66 \pm 0.16$               & $0.03 \pm 0.03$               & $2.52 \pm 0.34$           & $0.09 \pm 0.03$           & $20 \pm 3$        \\
\ttZ                            & $1.18 \pm 0.29$               & $1.05 \pm 0.25$               & $0.98 \pm 0.26$               & $0.76 \pm 0.27$               & $2.62 \pm 0.61$           & $1.57 \pm 0.32$           & $113 \pm 14$          \\
Others                          & $1.15 \pm 0.10$               & $0.76 \pm 0.11$               & $1.27 \pm 0.13$               & $0.74 \pm 0.07$               & $2.72 \pm 0.32$           & $1.64 \pm 0.15$           & $23 \pm 3$        \\
FNP                             & $1.04 \pm 0.37$               & $0.54 \pm 0.19$               & $0.62 \pm 0.47$               & $0.64 \pm 0.21$               & $1.68 \pm 0.46$           & $1.11 \pm 0.27$           & $7 \pm 1$       \\
\midrule
$\mu$                              & \multicolumn{2}{c}{$\mu_{\ttbar} = 1.11 \pm 0.08$}                          & \multicolumn{2}{c}{$\mu_{\ttbar} = 0.99 \pm 0.09$}                              & \multicolumn{2}{c}{$\mu_{\ttbar} = 1.10 \pm 0.09$}          & $\mu_{\ttZ} = 1.07 \pm 0.14$       \\
\bottomrule
\end{tabular}
\end{adjustbox}
\end{table}
\begin{figure}[htb!]
\begin{center}
\subfigure[]{
\includegraphics[width=0.48\textwidth]{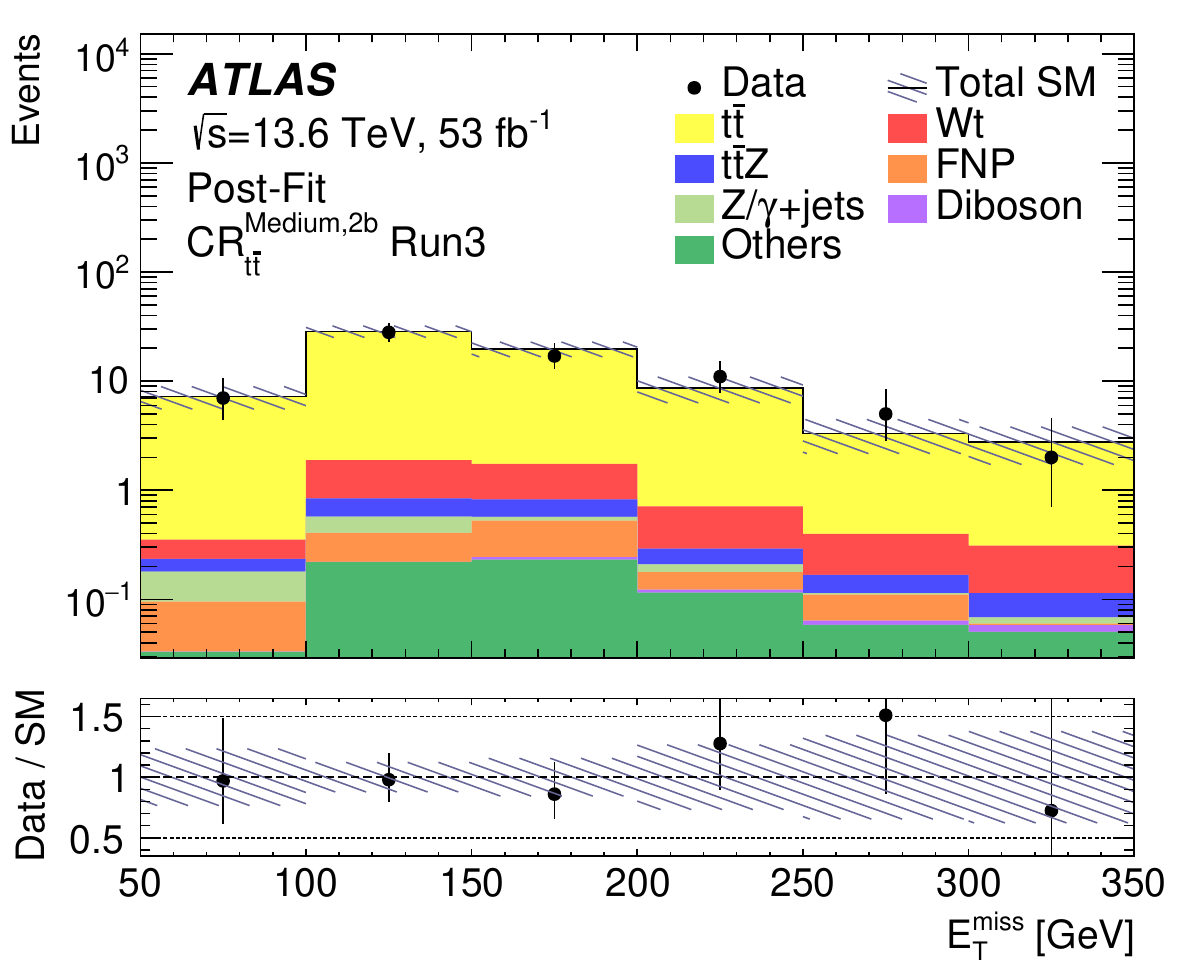}
}
\subfigure[]{
\includegraphics[width=0.48\textwidth]{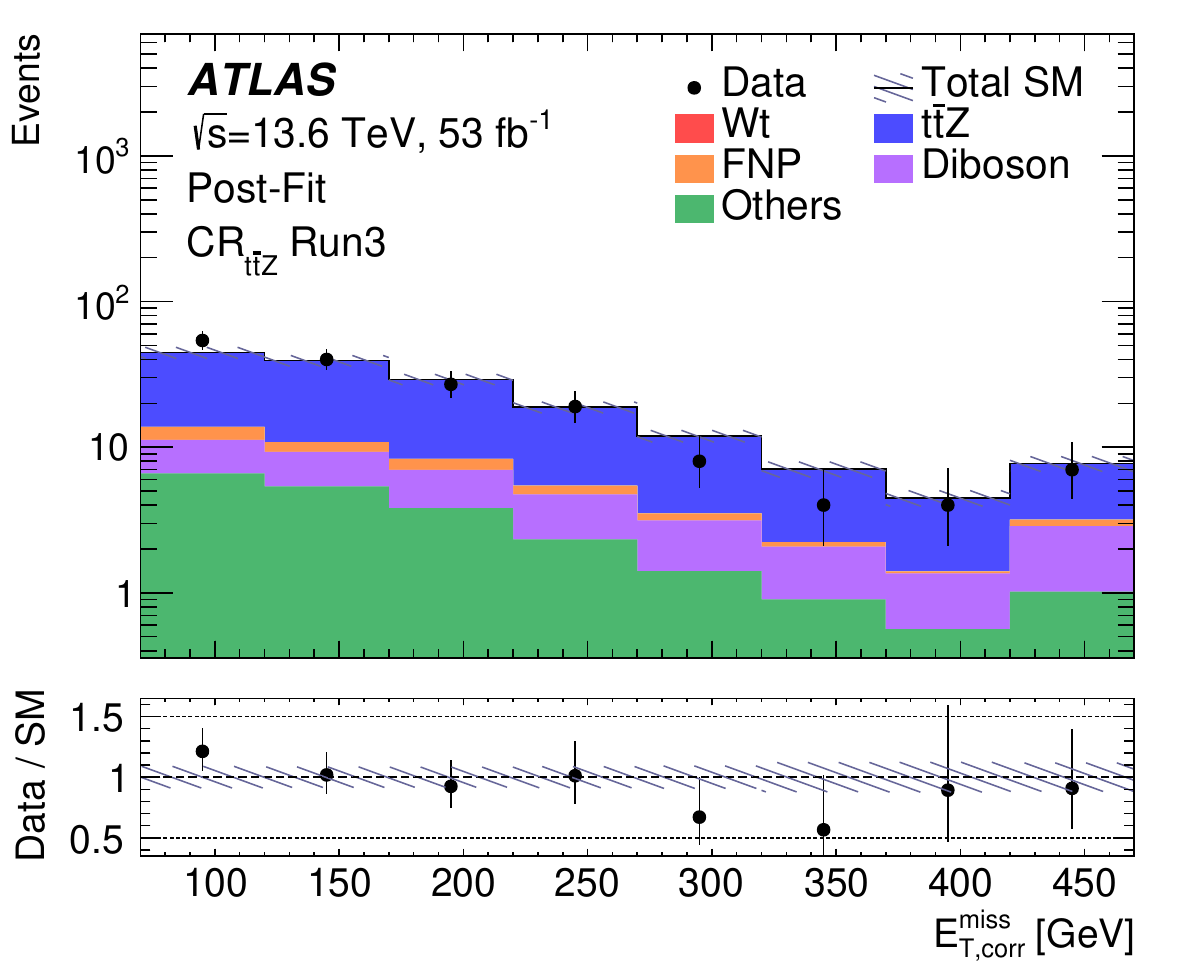}
}
\end{center}
\caption{Distributions of (a) the \MET in \CRttbar{Medium,2b}\ and (b) the \METcorr in \CRttZ, each after the background fit. The contributions from all SM backgrounds are shown as a histogram stack. ``Others'' includes  the contributions from $VVV$, $\ttbar t$, $\ttbar\ttbar$, $\ttbar W$, $\ttbar WW$, $\ttbar WZ$, $\ttbar H$, $tZ$ and $tWZ$ processes. The hatched bands represent the total combined statistical and systematic uncertainty. The rightmost bin includes overflow events. The bottom panels show the ratio of the observed data to the total SM background prediction, with hatched bands representing the total uncertainty of the background prediction.}
\label{fig:CRs}
\end{figure}
\begin{figure}[htb!]
\begin{center}
\subfigure[]{
\includegraphics[width=0.48\textwidth]{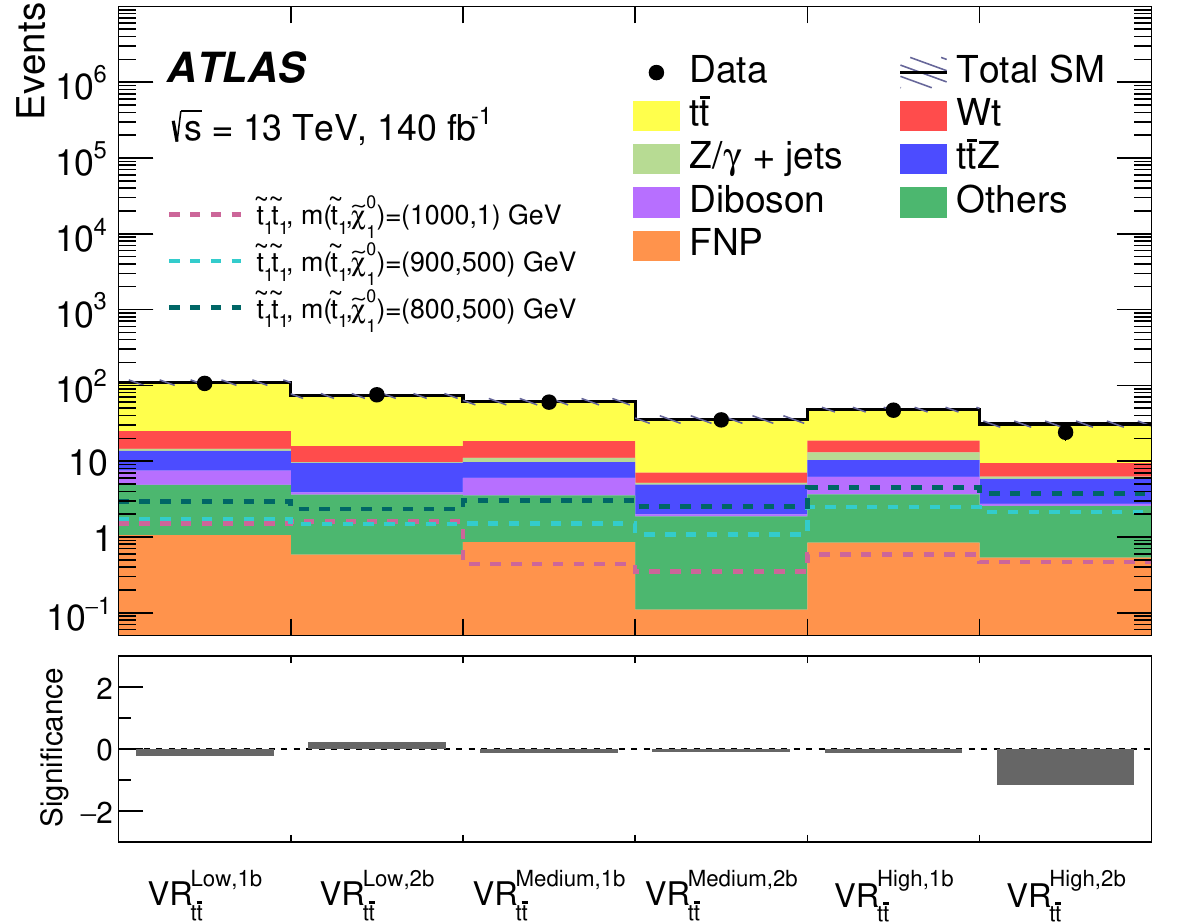}
}
\subfigure[]{
\includegraphics[width=0.48\textwidth]{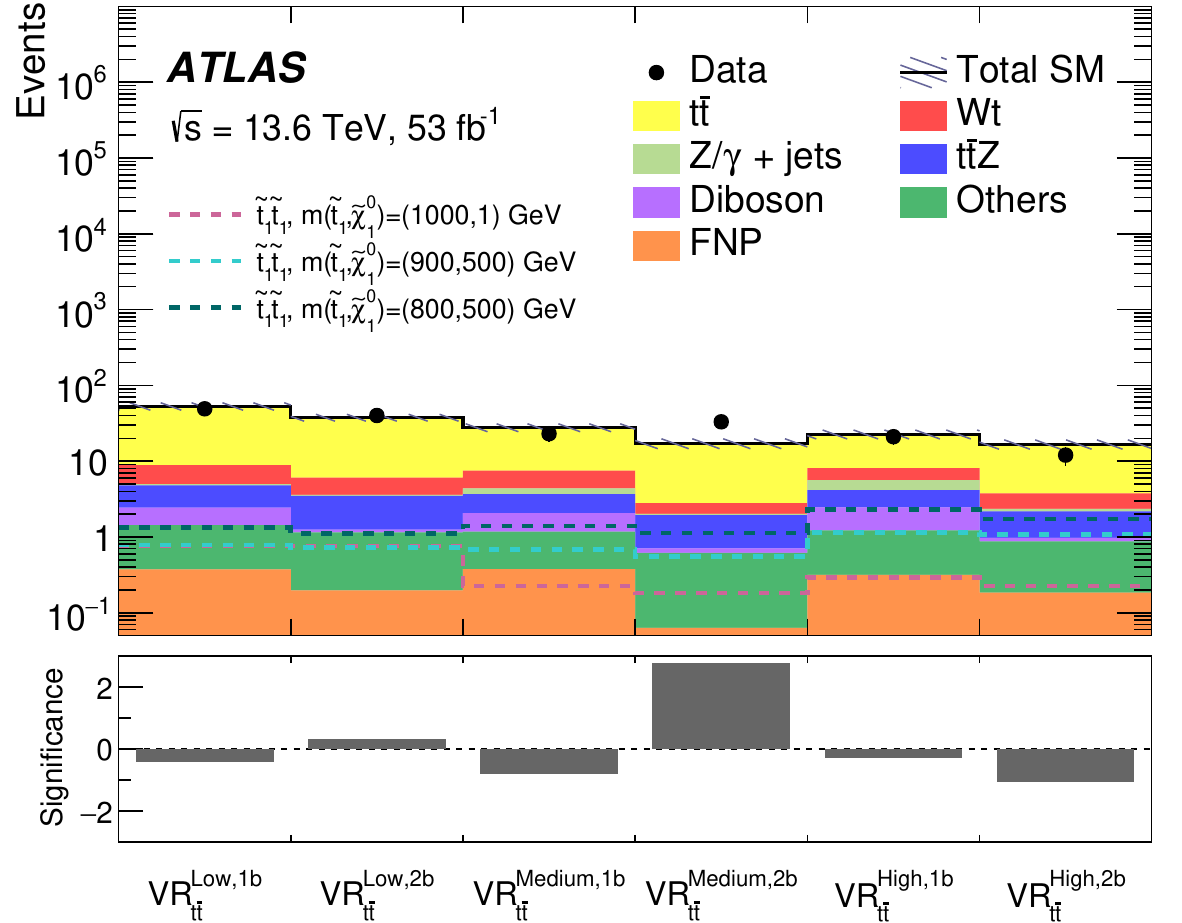}
}
\end{center}
\caption{Expected and observed yields in the \ttbar VRs for the (a) \RunTwo and (b) \RunThr datasets.
The upper panel shows the observed number of events together with the expected SM backgrounds obtained after the fit.
``Others'' includes  the contributions from $VVV$, $\ttbar t$, $\ttbar\ttbar$, $\ttbar W$, $\ttbar WW$, $\ttbar WZ$, $\ttbar H$, $tZ$ and $tWZ$ processes.
The hatched bands represent the total combined statistical and systematic uncertainty on the SM background.
The lower panel shows the significance as defined in Ref.~\cite{sig-atlas}.} %
\label{fig:VRpull}
\end{figure}

%


%
\FloatBarrier
\section{Systematic uncertainties}
\label{sec:systematics}

%

The uncertainty on the yield predictions in the analysis regions has roughly similar contributions from theoretical and modelling uncertainties of the simulated background, and the experimental limitations in the reconstruction of the physics objects.
The statistical uncertainty of the simulated events is also taken into account.

The main sources of experimental uncertainty are related to the jet energy scale (JES) and the jet energy resolution (JER).
The JES and JER uncertainties are derived as a function of the \pT and $\eta$ of the jet, as well as of the \pileup conditions and the jet-flavour composition of the selected jet sample, and are evaluated using a series of simulation-based techniques and in situ measurements, documented in Ref.~\cite{JETM-2018-05}.
Uncertainties associated with the modelling of the $b$-tagging efficiencies for $b$-jets, $c$-jets and light-flavour jets are also considered~\cite{FTAG-2019-02, FTAG-2020-08,FTAG-2018-01}. %
The systematic uncertainties related to the modelling of the \MET are estimated by propagating the uncertainties in the energy and momentum scales of electrons, muons, and jets, and the uncertainties in the resolution and scale of the soft term~\cite{JETM-2020-03}.
Other detector-related systematic uncertainties, including those arising from lepton reconstruction efficiency, energy scale, energy resolution and in the modelling of the trigger efficiency~\cite{EGAM-2018-01,EGAM-2021-02,MUON-2022-01,TRIG-2018-05,TRIG-2018-01}, together with the ones due to the \pileup reweighting~\cite{SIMU-2020-01} and JVT~\cite{ATLAS-CONF-2014-018} are also considered, and found to have a small impact on the results.
Finally, the uncertainty on the full Run 2 integrated luminosity amounts to $0.83\%$~\cite{DAPR-2021-01}, while the uncertainty on the combined 2022--2023 integrated luminosity is estimated to be $2\%$~\cite{ATL-DAPR-PUB-2024-001}, and is uncorrelated with that of Run 2.

The uncertainty on the modelling of the $\ttbar$ background due to initial-state radiation (ISR) was estimated by simultaneously varying the parameter that regulates the high-\pT radiation against which the \ttbar system recoils, and the normalisation and factorisation scales~\cite{ATL-PHYS-PUB-2016-020}.
Variations of the strong coupling constant $\alpha_S$ applied to the modelling of initial-state radiation are evaluated as described in Ref.~\cite{ATL-PHYS-PUB-2017-007}.
The impact of final-state radiation (FSR) was evaluated by varying the renormalisation scale for emissions from the parton shower up or down by a factor two.
The uncertainties on $\ttZ$ production also include the effects of QCD scale uncertainties. Additionally, the uncertainties due to possible additional ISR contributions are evaluated by varying the \texttt{Var3c} parameter in the \texttt{A14} \PYTHIA tune~\cite{Sjostrand:2014zea}.
For both \ttbar and \ttZ production, the impact of using a different parton shower and hadronisation model was evaluated
by comparing simulated samples generated using the two different showering models implemented in \PYTHIA and in \HERWIG~\cite{Bahr:2008pv,Bellm:2015jjp}.
For single top production in the $Wt$ channel, an uncertainty is assigned to account for the interference with \ttbar production, which is estimated by comparing samples generated using the diagram removal (nominal) and the diagram subtraction prescriptions~\cite{Frixione:2008yi}.
For all the processes mentioned above, the PDF uncertainties~\cite{Butterworth:2015oua} were evaluated and found to be negligible.
For minor backgrounds, the uncertainty on the cross section used for the normalisation is included.

A summary of the impact of the systematic uncertainties on the background expectation in each SR bin is presented in~\cref{fig:SystBreakdown}.
The total systematic uncertainty in the SR bins ranges between 10\% and 23\% for \RunTwo samples, and between 14\% and 29\% for \RunThr samples. The dominant sources of uncertainty are the JES and JER uncertainties, the theoretical uncertainties, the MC statistical uncertainty and the uncertainty in the background normalisation.
All systematic uncertainties are considered uncorrelated between Run 2 and Run 3, and this choice is not found to impact the final results.

\begin{figure}[htb!]
\begin{center}
\subfigure[]{
\includegraphics[width=0.98\textwidth]{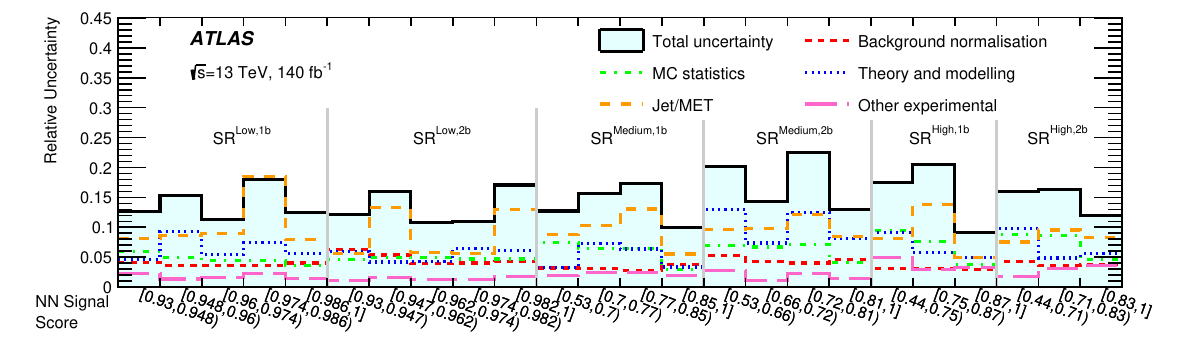}
}
\subfigure[]{
\includegraphics[width=0.98\textwidth]{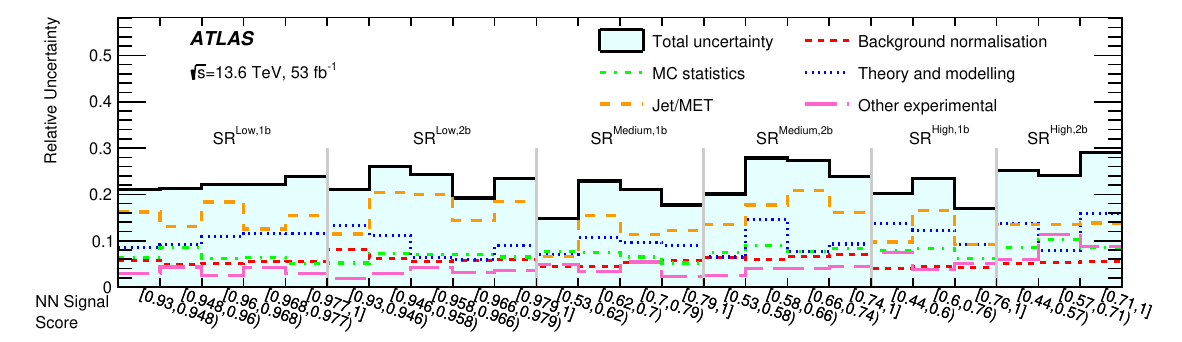}
}
\end{center}
\caption{Comparison of the relative uncertainty for the total background yield in each SR for the (a) \RunTwo and (b) \RunThr datasets, including the contribution from the different sources of uncertainty, calculated as results of the profile likelihood fit.
The category ``Other experimental'' includes minor detector uncertainties, such as the ones related to leptons, flavour tagging and \pileup modelling. The total uncertainty is calculated considering the correlation across the various systematic components, and therefore it does not necessarily correspond to the sum of the single components.}
\label{fig:SystBreakdown}
\end{figure}

The SUSY signal cross section uncertainty is evaluated from an envelope of the cross section predictions using different PDF sets and factorisation and renormalisation scales as described in Ref.~\cite{Borschensky:2014cia,Beenakker:2024nlf}.
The SUSY theory uncertainties affecting the signal acceptance in the SRs are computed from the variation of the renormalisation and factorisation scales. Additionally, an uncertainty of 10\% is added in the SRs to take into account the uncertainties due to the  variation of the fragmentation and merging scales.
This uncertainty is estimated with alternative samples for selected benchmark signal models.


%
\FloatBarrier
\section{Results and interpretation}
%

The observed data and the predicted SM event yields obtained from a CR-only fit are shown in \cref{fig:SRpull} for all the SR bins and separately for \RunTwo and \RunThr datasets. The selections for the three categories (Low, Medium, and High) are not mutually exclusive, as for their definition in \ref{sec:analysis-strategy}.
Furthermore, the expected yields of representative benchmark signal models and the signal significance of the observed data to the post-fit SM expectation, calculated as described in Ref.~\cite{ATL-PHYS-PUB-2020-025}, are also shown.
The SM predictions show a good agreement with the observed data within the corresponding uncertainties: no significant excess above the SM expectation is observed in data and the largest fluctuations (observed in \RunTwo \(\text{SR}^{\text{Medium,1b}}\) and \(\text{SR}^{\text{High,1b}}\)) are below two standard deviations.
\begin{figure}[htb!]
\begin{center}
\subfigure[]{
\includegraphics[width=0.98\textwidth]{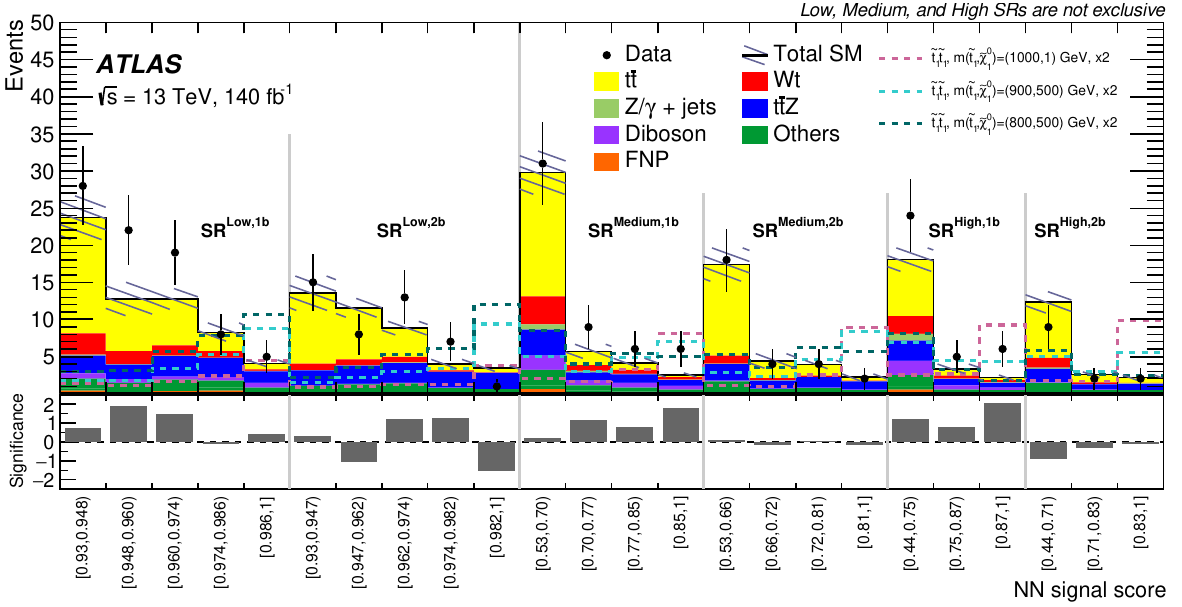}
}
\subfigure[]{
\includegraphics[width=0.98\textwidth]{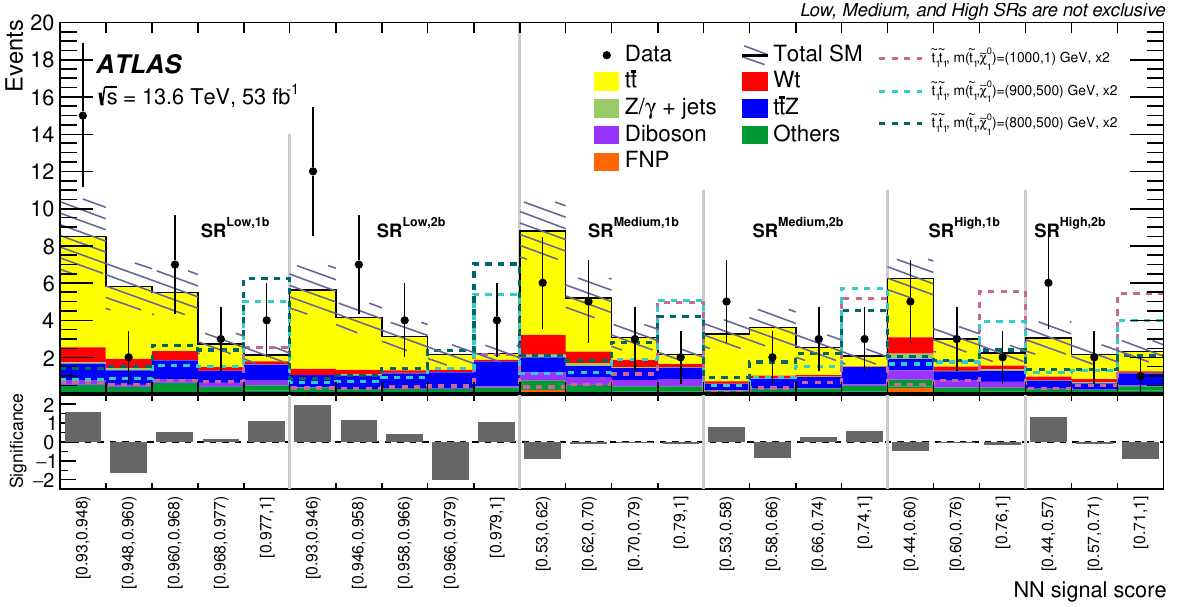}
}
\end{center}
\caption{Expected and observed yields in the SR bins for the (a) \RunTwo and (b) \RunThr datasets.
The upper panel shows the observed number of events together with the expected SM backgrounds obtained after the fit. The normalisation of the signal samples is increased by a factor of two for visualisation purposes.
``Others'' includes  the contributions from $VVV$, $\ttbar t$, $\ttbar\ttbar$, $\ttbar W$, $\ttbar WW$, $\ttbar WZ$, $\ttbar H$, $tZ$ and $tWZ$ processes.
The shaded band represents the total uncertainty in the expected SM background.
The lower panel shows the significance as defined in~\cite{sig-atlas}.}
\label{fig:SRpull}
\end{figure}

The analysis results are interpreted in terms of both model-dependent exclusion limits and model-independent upper limits on new physics cross section.
Model-dependent exclusion limits are derived for the signal scenarios considered in the analysis.
As a first step, separate model-dependent limits are derived using only the regions from either the low, the medium or the high categories. For each combination of SUSY particle masses,
only the category with the best expected sensitivity is then considered for the final limit setting. The hypothesis tests include the expected signal yields and their associated uncertainties in both the CRs and SRs. All exclusion limits are quoted at 95\% confidence level (CL) using the CL$_\mathrm{s}$ method~\cite{Read:2002hq}.
When setting limits, CRs and the SR bins within each \DeltaM-category are statistically combined.
Limits for simplified models in which pair-produced top squarks decay with 100\% branching ratio into a top quark and the lightest neutralino are presented in the \stopone--\ninoone mass plane in \cref{fig:excl_bestexp_combined}.
Top-squark masses up to $1060~\GeV$ are excluded for a massless lightest neutralino, improving the \RunTwo results by about 100 \GeV, while neutralino masses up to $560~\GeV$ are excluded for $m(\stopone)$ values of 900~\GeV. Observed limits are weaker than the expected limits, due to the small excess observed in the \(1b\) regions in \RunTwo (medium and high).
The analysis was tested also for signal models in which the mass difference between the \stopone and the \ninoone is not enough to allow the production of on-shell top quarks, resulting in the following 3-body decay: \(\stopone \rightarrow bW\ninoone\). Some sensitivity for these models is observed, especially close to the kinematic edge, even if the NN discriminant was not specifically trained for these models.
The observed limits are not continuous, since for each combination of sparticle masses, the \DeltaM\ category with best expected sensitivity is considered for the final limit setting. The
change in best expected category is responsible for the discontinuity in the limits.
\cref{fig:excl_bestexp_combined} also includes, with a green dashed line, the expected evidence sensitivity for this analysis obtained by signal injection, in case a significant excess with respect to the SM prediction would have been observed.
\begin{figure}[!ht] %
\centering
\subfigure{\includegraphics[width=0.6\linewidth]{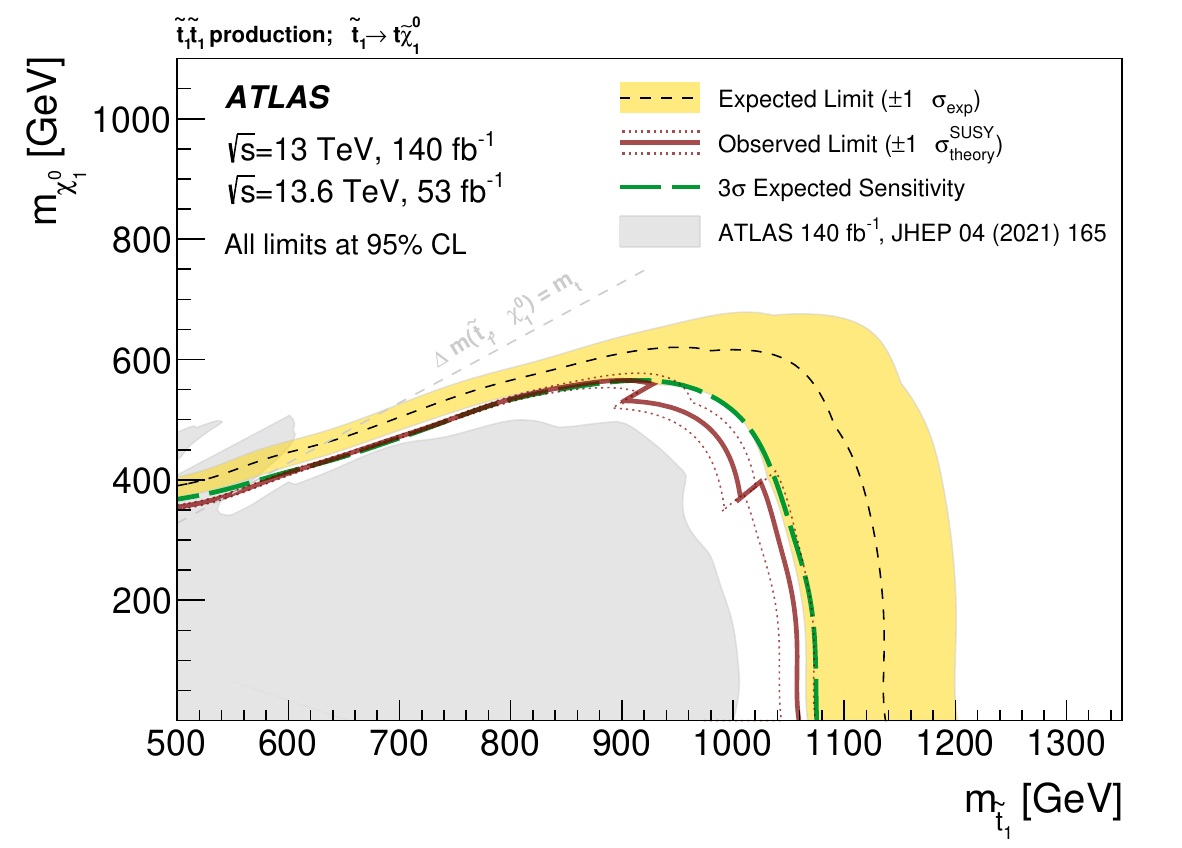}}
\caption{Expected $95\%$ CL limits (dashed line), with their $\pm1\sigma$ uncertainty (yellow band), and observed limits (solid red line) on the $\stopone$-$\ninoone$ mass plane for the low-, medium-, and high-\deltam regions combined (best-expected CL$_s$), combining \RunTwo and \RunThr. The green dashed line represents the sensitivity of this analysis for a \(3\sigma\) evidence.
The shaded region represents the observed exclusion limits of the previous full-\RunTwo analysis~\cite{SUSY-2018-08}. The dotted lines show the effect on the observed limit when varying the signal cross section by $\pm1\sigma$ of the theoretical uncertainty.}
\label{fig:excl_bestexp_combined}
\end{figure}
\\
Model-independent upper limits on the visible cross section $(\sigma_{\mathrm{vis}})$ of new physics are defined as the 95\% CL upper limit on the number of signal events, $(S^{95})$, divided by the integrated luminosity.
These limits are derived using single-bin SRs constructed by merging the three highest NN signal score bins for each SR, referred to as SRdisc.
In each SRdisc, the upper limit on $\sigma_{\mathrm{vis}}$ is derived by performing a model-independent hypothesis test~\cite{Baak:2014wma}, in which a free signal component is introduced as an additional process to be constrained by the observed yield, assuming negligible signal contamination in the CRs.  %
The resulting model-independent upper limits on $\sigma_{\mathrm{vis}}$ are presented in Table~\ref{tab:ULMI}.
\begin{table}[!ht]
\begin{center}
\caption{Left to right: number of observed and expected events in the discovery SR (second and third column),
95\%~CL upper limits on the visible cross section ($\sigma_{\mathrm{vis}}$) and on the number of signal events ($S_{\mathrm{obs}}^{95}$ ).
The sixth  column ($S_{\mathrm{exp}}^{95}$) shows the 95\% CL upper limit on the number of signal events, given the expected number (and $\pm 1\sigma$ variation on the expectation) of background events.
In the last two columns, CL$_\mathrm{B}$ provides a measure of the compatibility of the observed data with the signal strength hypothesis at the 95\% CL limit relative to fluctuations of the background, and the discovery $p$-value ($p(s = 0)$) measures compatibility of the observed data with the background-only hypothesis relative to fluctuations of the background.
The $p$-value is reported as $0.5$ if the observed yield is smaller than that predicted.}
\label{tab:ULMI}
\renewcommand{\arraystretch}{1.2}
\setlength{\tabcolsep}{0.0pc}{\footnotesize
\begin{tabular*}{\textwidth}{@{\extracolsep{\fill}}lcccccccc}
\toprule{Discovery SR}   & Observed & Expected        & $\sigma_{\mathrm{vis}}$ [fb] & $S_{\mathrm{obs}}^{95}$ & $S_{\mathrm{exp}}^{95}$                & CL$_\mathrm{B}$ & $p(s=0)$ ($Z$)  \\
\midrule
SRdisc$^{\mathrm{Low,1b}}$ (\RunTwo)    & 32       & $26.4\pm 2.4$      & $0.12$    &  $16.8$   & $ { 13 }^{ +6 }_{ -4 }$   & $0.78$    & $ 0.20$~$(0.82)$ \\%
SRdisc$^{\mathrm{Low,2b}}$ (\RunTwo)    & 21       & $17.7\pm 2.0$      & $0.09$    &  $13.2$   & $ { 10.4 }^{ +4.7 }_{ -3.2 }$   & $0.74$    & $ 0.25$~$(0.68)$ \\%
SRdisc$^{\mathrm{Medium,1b}}$ (\RunTwo) & 21       & $13.2\pm 1.3$      & $0.12$    &  $16.3$   & $ {  9.0 }^{ +4.2 }_{ -2.8 }$   & $0.94$    & $ 0.04$~$(1.75)$ \\%
SRdisc$^{\mathrm{Medium,2b}}$ (\RunTwo) & 10       & $10.3\pm 1.4$      & $0.05$    &  $7.7$    & $ {  8.0 }^{ +3.8 }_{ -2.5 }$   & $0.45$    & $ 0.50$~$(0.00)$ \\%
SRdisc$^{\mathrm{High,1b}}$ (\RunTwo)   & 35       & $25.0\pm 1.8$      & $0.15$    &  $20.8$   & $ { 12 }^{ +5 }_{ -4 }$   & $0.94$    & $ 0.05$~$(1.69)$ \\%
SRdisc$^{\mathrm{High,2b}}$ (\RunTwo)   & 13       & $15.7\pm 1.7$      & $0.05$    &  $7.5$    & $ { 9.5 }^{ +4.4 }_{ -2.9 }$   & $0.26$    & $ 0.50$~$(0.00)$ \\%
\midrule
SRdisc$^{\mathrm{Low,1b}}$ (\RunThr)    & 14       & $10.6\pm 1.2$      & $0.21$    &  $11.0$   & $ {  8.1 }^{ +3.9 }_{ -2.5 }$   & $0.77$    & $ 0.21$~$(0.82)$ \\%
SRdisc$^{\mathrm{Low,2b}}$ (\RunThr)    & 8        & $7.7\pm 1.1$       & $0.14$    &  $7.3$    & $ {  7.1 }^{ +3.5 }_{ -2.3 }$   & $0.52$    & $ 0.47$~$(0.07)$ \\%
SRdisc$^{\mathrm{Medium,1b}}$ (\RunThr) & 10       & $10.1\pm 1.1$      & $0.14$    &  $7.6$    & $ {  7.9 }^{ +3.8 }_{ -2.5 }$   & $0.46$    & $ 0.50$~$(0.00)$ \\%
SRdisc$^{\mathrm{Medium,2b}}$ (\RunThr) & 8        & $8.1\pm 1.1$       & $0.13$    &  $7.1$    & $ {  7.2 }^{ +3.5 }_{ -2.5 }$   & $0.48$    & $ 0.50$~$(0.00)$ \\%
SRdisc$^{\mathrm{High,1b}}$ (\RunThr)   & 10       & $11.2\pm 1.2$      & $0.13$    &  $7.1$    & $ {  8.2 }^{ +3.9 }_{ -2.5 }$   & $0.36$    & $ 0.50$~$(0.00)$ \\%
SRdisc$^{\mathrm{High,2b}}$ (\RunThr)   & 9        & $7.6\pm 1.0$       & $0.16$    &  $8.3$    & $ {  7.1 }^{ +3.5 }_{ -2 .2}$   & $0.65$    & $ 0.33$~$(0.45)$ \\%
\bottomrule
\end{tabular*}
}
\end{center}
\end{table}
The \mttwoL distributions for the predicted and observed events, in each SRdisc defined for the \RunThr dataset are then shown in \cref{fig:SRdisc_distrib}.
\begin{figure}[htb!]
\begin{center}
\subfigure[]{\includegraphics[width=0.32\textwidth]{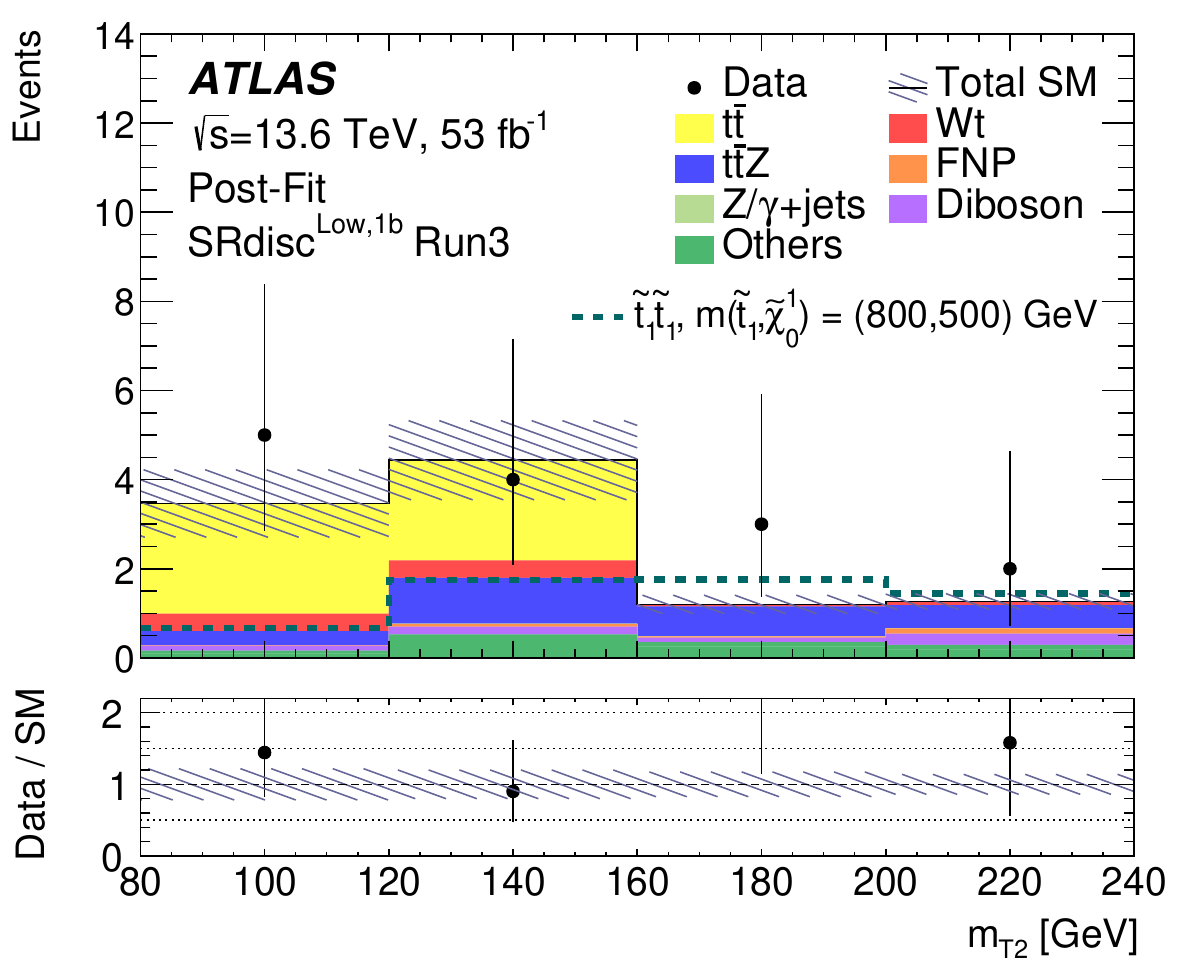}}
\subfigure[]{\includegraphics[width=0.32\textwidth]{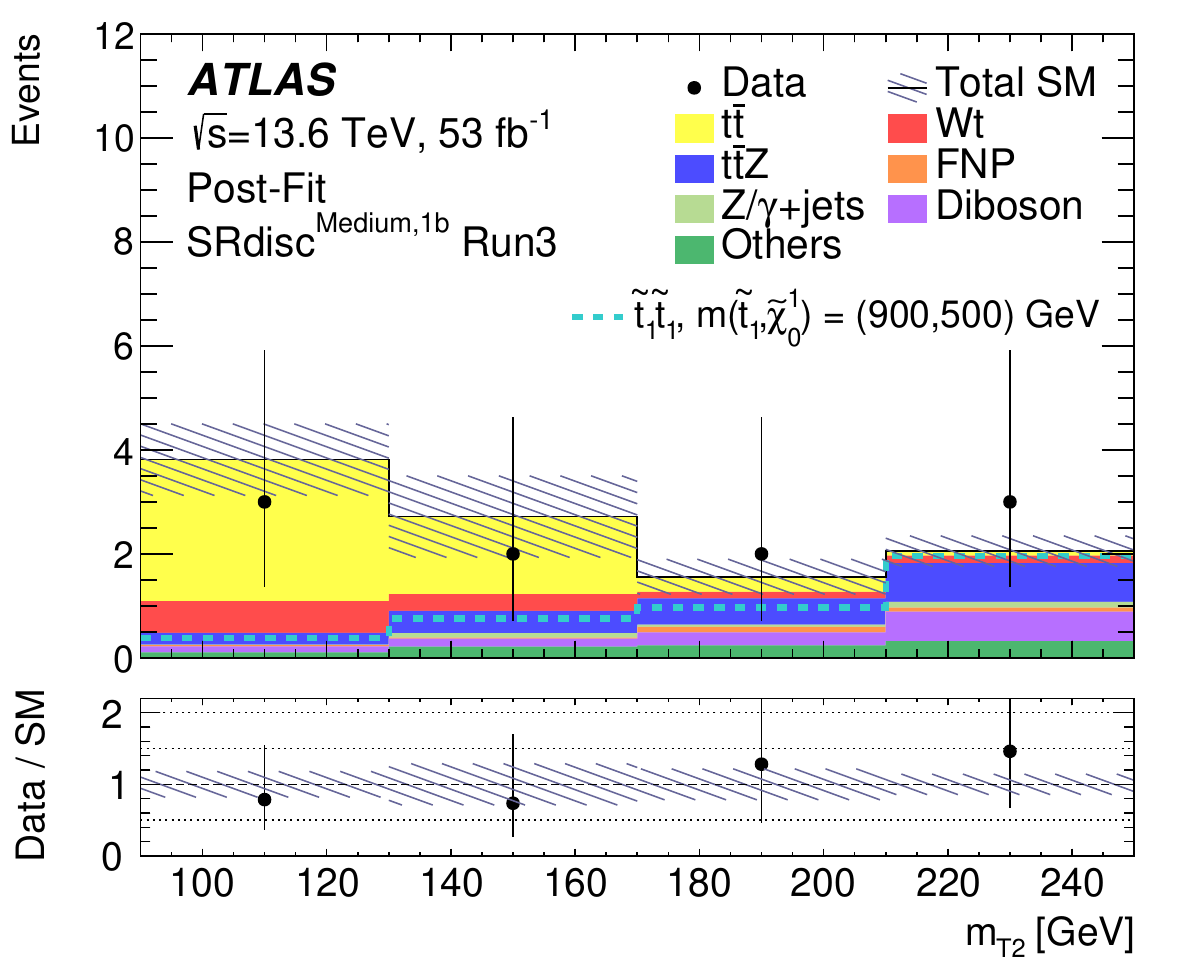}}
\subfigure[]{\includegraphics[width=0.32\textwidth]{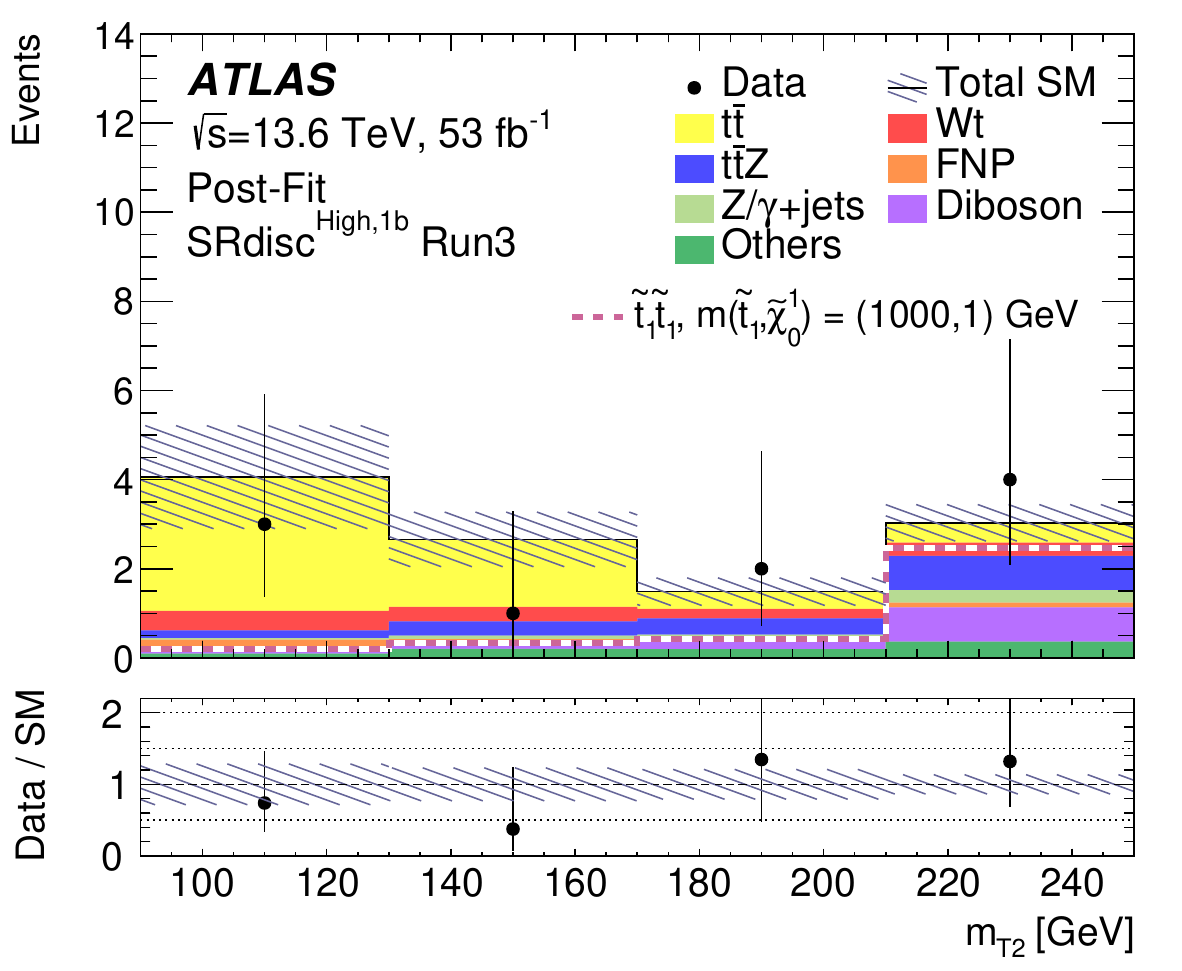}} \\
\subfigure[]{\includegraphics[width=0.32\textwidth]{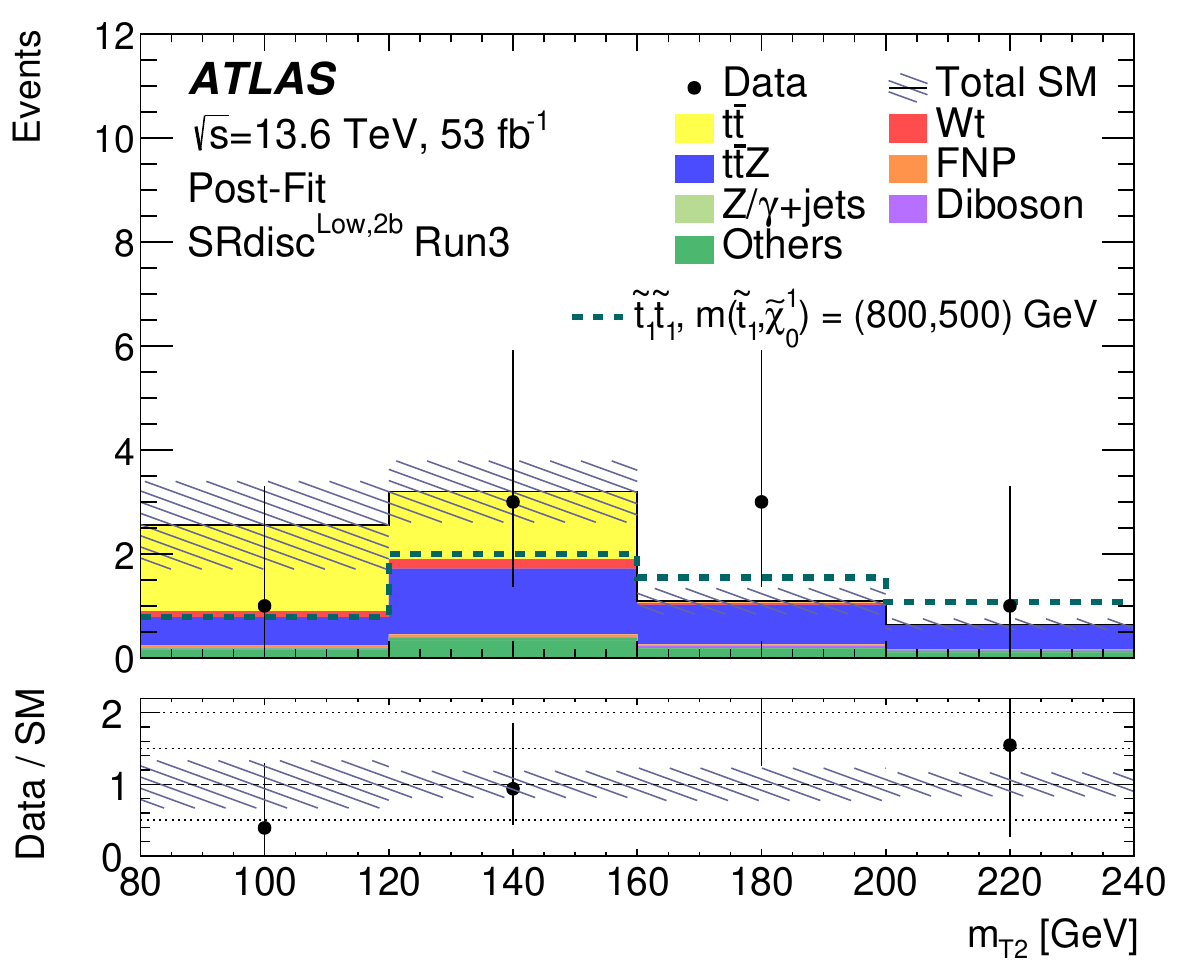}}
\subfigure[]{\includegraphics[width=0.32\textwidth]{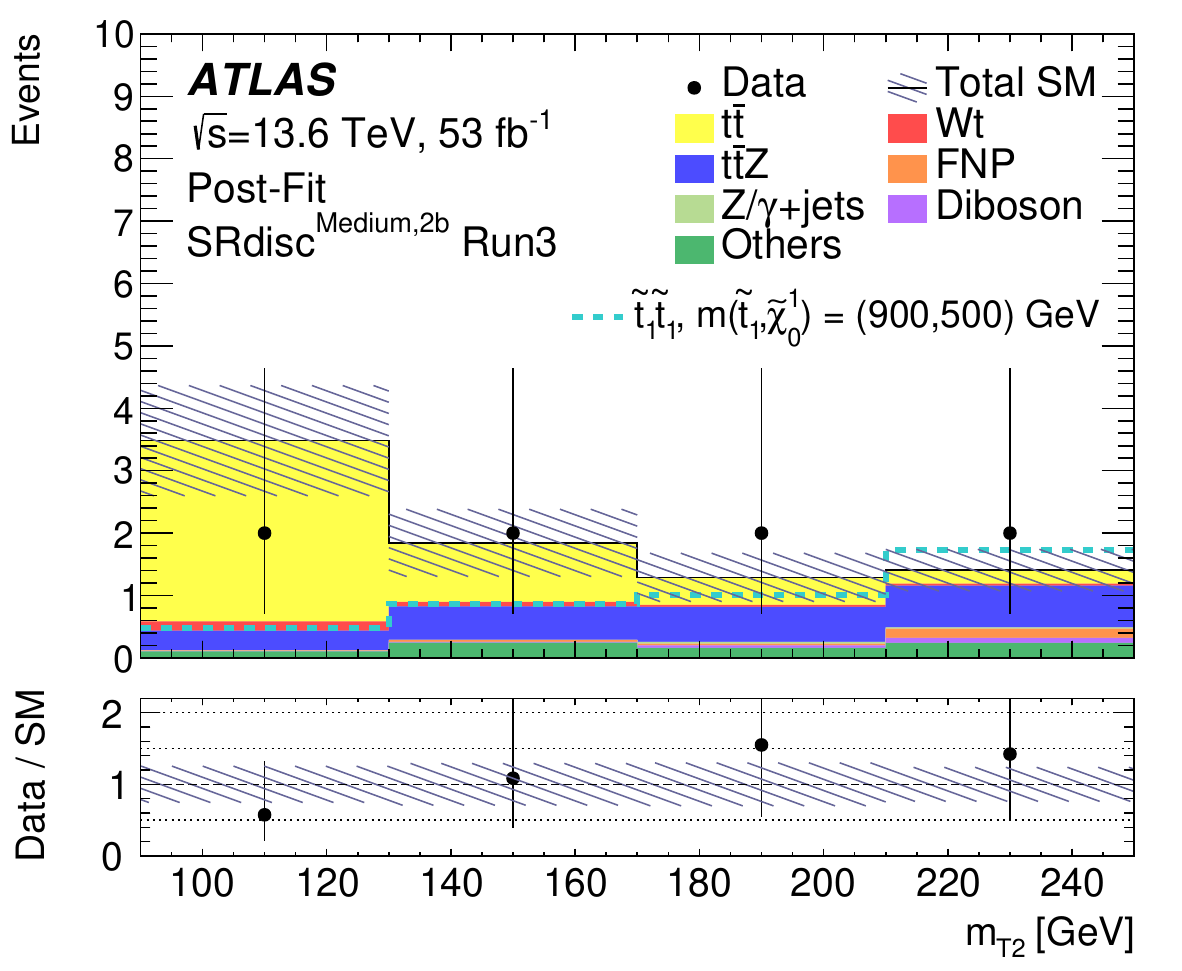}}
\subfigure[]{\includegraphics[width=0.32\textwidth]{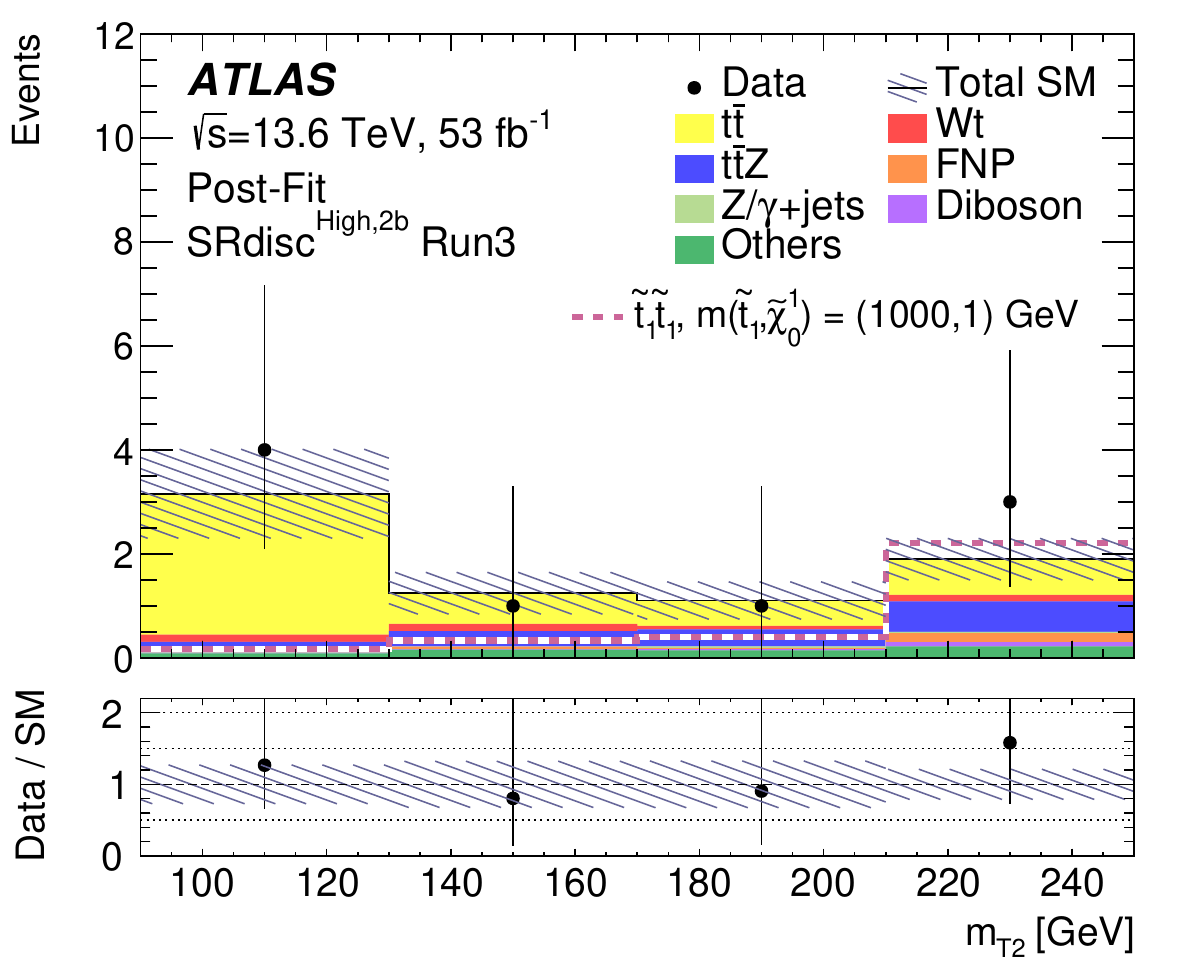}}
\end{center}
\caption{Expected and observed distributions of \mttwoL in each \RunThr discovery region SRdisc, in events with one $b$-tagged jet (first row), and two $b$-tagged jets (second row) for the low-\DeltaM\ (left), medium-\DeltaM\ (centre) and high-\DeltaM\ (right) regions.
The contributions from all SM backgrounds are shown as a histogram stack. ``Others'' includes  the contributions from $VVV$, $\ttbar t$, $\ttbar\ttbar$, $\ttbar W$, $\ttbar WW$, $\ttbar WZ$, $\ttbar H$, $tZ$ and $tWZ$ processes.
The hatched bands represent the total combined statistical and systematic uncertainty. The rightmost bin includes overflow events. Reference signal models are overlaid for comparison: $m(\stopone,\ninoone) = (800, 500)~\GeV$ for the low-\DeltaM\ regions, $m(\stopone,\ninoone) = (900, 500)~\GeV$ for the medium, and $m(\stopone,\ninoone) = (1000, 1)~\GeV$ for the high.
The bottom panels show the ratio of the observed data to the total SM background prediction, with hatched bands representing the total uncertainty of the background prediction.}
\label{fig:SRdisc_distrib}
\end{figure}


%
\FloatBarrier
\section{Conclusions}

This paper presents a search for the direct top squark pair production, with each top squark decaying directly into a top quark and the lightest neutralino, $\stopone\to~t\ninoone$.
The analysis targets final states containing exactly two oppositely charged leptons, \(b\)-tagged jets and missing transverse momentum.
The search uses the full LHC Run~2 dataset, corresponding to an integrated luminosity of $140~\ifb$ of \(pp\) collisions at $\sqrt{s}=13~\TeV$ collected by the ATLAS experiment between 2015 and 2018, and $53~\ifb$ of \(pp\) collisions at $\sqrt{s}=13.6~\TeV$ collected in 2022 and 2023 during the Run~3 of the LHC.

Compared to previous searches, a significant improvement in the expected sensitivity, especially at high \ninoone masses, is obtained through the inclusion of the additional integrated luminosity, the improvement in the event reconstruction techniques, and the development of a new analysis strategy based on machine-learning techniques. For a benchmark signal point, considering a $1~\TeV$ \stopone and a $400~\GeV$ \ninoone, the expected improvement of the upper limit on the signal strength arising from the improved reconstruction and new analysis strategy is 50\%. A further 20\% is then reached thanks to the addition of Run 3 data. These improvements allow the dilepton channel to be competitive in terms of sensitivity with the single-lepton analysis,
despite the latter having an advantage in the branching fraction.

The observed data are found to be consistent with the Standard Model expectations. Exclusion limits at the 95\% confidence level are set on \stopone and \ninoone masses, up to approximately $1060~\GeV$ and  $560~\GeV$, respectively. %
These limits are slightly weaker than expected due to the presence of small excesses of data over the predicted background, compatible with statistical fluctuations.
The analysis extends the previous ATLAS limits obtained in the dilepton channel by \(\approx10\%\) in mass reach.

\clearpage

\section*{Acknowledgements}
%

%

%
%

%
%

We thank CERN for the very successful operation of the LHC and its injectors, as well as the support staff at
CERN and at our institutions worldwide without whom ATLAS could not be operated efficiently.

The crucial computing support from all WLCG partners is acknowledged gratefully, in particular from CERN, the ATLAS Tier-1 facilities at TRIUMF/SFU (Canada), NDGF (Denmark, Norway, Sweden), CC-IN2P3 (France), KIT/GridKA (Germany), INFN-CNAF (Italy), NL-T1 (Netherlands), PIC (Spain), RAL (UK) and BNL (USA), the Tier-2 facilities worldwide and large non-WLCG resource providers. Major contributors of computing resources are listed in Ref.~\cite{ATL-SOFT-PUB-2026-001}.

We gratefully acknowledge the support of ANPCyT, Argentina; YerPhI, Armenia; ARC, Australia; BMWFW and FWF, Austria; ANAS, Azerbaijan; CNPq and FAPESP, Brazil; NSERC, NRC and CFI, Canada; CERN; ANID, Chile; CAS, MOST and NSFC, China; Minciencias, Colombia; MEYS CR, Czech Republic; DNRF and DNSRC, Denmark; IN2P3-CNRS and CEA-DRF/IRFU, France; SRNSFG, Georgia; BMFTR, HGF and MPG, Germany; GSRI, Greece; RGC and Hong Kong SAR, China; ICHEP and Academy of Sciences and Humanities, Israel; INFN, Italy; MEXT and JSPS, Japan; CNRST, Morocco; NWO, Netherlands; RCN, Norway; MNiSW, Poland; FCT, Portugal; MNE/IFA, Romania; MSTDI, Serbia; MSSR, Slovakia; ARIS and MVZI, Slovenia; DSI/NRF, South Africa; MICIU/AEI, Spain; SRC and Wallenberg Foundation, Sweden; SERI, SNSF and Cantons of Bern and Geneva, Switzerland; NSTC, Taipei; TENMAK, T\"urkiye; STFC/UKRI, United Kingdom; DOE and NSF, United States of America.

Individual groups and members have received support from BCKDF, CANARIE, CRC and DRAC, Canada; CERN-CZ, FORTE and PRIMUS, Czech Republic; COST, ERC, ERDF, Horizon 2020 and Marie Sk{\l}odowska-Curie Actions, European Union; Investissements d'Avenir Labex, Investissements d'Avenir Idex and ANR, France; DFG and AvH Foundation, Germany; Herakleitos, Thales and Aristeia programmes co-financed by EU-ESF and the Greek NSRF, Greece; BSF-NSF and MINERVA, Israel; NCN and NAWA, Poland; La Caixa Banking Foundation, CERCA and AGAUR programs from Generalitat de Catalunya and PROMETEO and GenT Programmes Generalitat Valenciana, Spain; G\"{o}ran Gustafssons Stiftelse, Sweden; The Royal Society and Leverhulme Trust, United Kingdom; Eric and Wendy Schmidt Fund for Strategic Innovation, United States of America.

In addition, individual members wish to acknowledge support from Chile: Agencia Nacional de Investigaci\'on y Desarrollo (ANID FONDECYT reg. 1230987, FONDECYT 1230812, FONDECYT 1240864, Fondecyt 3240661, Fondecyt Regular 1240721); China: Chinese Ministry of Science and Technology (MOST-2023YFA1605700, MOST-2023YFA1609300), National Natural Science Foundation of China (NSFC 12275265, NSFC-W2543005); Czech Republic: Czech Science Foundation (GACR - 24-11373S), Ministry of Education Youth and Sports (ERC-CZ-LL2327, FORTE CZ.02.01.01/00/22\_008/0004632), PRIMUS Research Programme (PRIMUS/21/SCI/017); EU: H2020 European Research Council (ERC - 101002463); European Union: European Research Council (BARD No. 101116429, ERC - 948254, ERC 101089007), European Regional Development Fund (HE COFUND GA No.101081355, ERDF), Marie Sklodowska-Curie Actions (GAP-101168829); France: Agence Nationale de la Recherche (ANR-21-CE31-0013, ANR-22-EDIR-0002, ANR-24-CE31-0504-01); Germany: Deutsche Forschungsgemeinschaft (DFG - 469666862); China: Research Grants Council (GRF); Italy: Ministero dell'Università e della Ricerca (NextGenEU 153D23001490006 M4C2.1.1, NextGenEU I53D23000820006 M4C2.1.1, NextGenEU I53D23001490006 M4C2.1.1, SOE2024\_0000023); Japan: Japan Society for the Promotion of Science (JSPS KAKENHI  JP25H0063, JSPS KAKENHI JP22H01227, JSPS KAKENHI JP22H04944, JSPS KAKENHI JP22KK0227, JSPS KAKENHI JP24K23939, JSPS KAKENHI JP24KK0251, JSPS KAKENHI JP25H00650, JSPS KAKENHI JP25H01291, JSPS KAKENHI JP25K01023); Poland: Polish National Science Centre (NCN 2021/42/E/ST2/00350, NCN OPUS 2023/51/B/ST2/02507, NCN OPUS nr 2022/47/B/ST2/03059, NCN UMO-2019/34/E/ST2/00393, UMO-2022/47/O/ST2/00148, UMO-2023/49/B/ST2/04085, UMO-2023/51/B/ST2/00920, UMO-2024/53/N/ST2/00869); Spain: Agència de Gestió d¿Ajuts Universitaris i de Recerca. (AGAUR - 2023 BP 00141), Ministry of Science and Innovation (RYC2019-028510-I, RYC2020-030254-I, RYC2021-031273-I, RYC2022-038164-I), Ministerio de Ciencia, Innovación y Universidades/Agencia Estatal de Investigaci\'on (PID2022-142604OB-C22); Sweden: Carl Trygger Foundation (Carl Trygger Foundation CTS 22:2312), Swedish Research Council (Swedish Research Council 2023-04654, VR 2021-03651, VR 2022-03845, VR 2022-04683, VR 2023-03403, VR 2024-05451, VR 2025-05940), Knut and Alice Wallenberg Foundation (KAW 2023.0366); Switzerland: Swiss National Science Foundation (SNSF - PCEFP2\_194658); United Kingdom: The Binks Trust, Royal Society (NIF-R1-231091); United States of America: U.S. Department of Energy (ECA DE-AC02-76SF00515), John Templeton Foundation (John Templeton Foundation 63206), Neubauer Family Foundation.

%
%


%
\clearpage

\printbibliography
\clearpage
\input{atlas_authlist}
%

%
%

%

%
%
%
%

\end{document}

%% file: atlas_authlist.tex
 
\begin{flushleft}
\hypersetup{urlcolor=black}
{\Large The ATLAS Collaboration}

\bigskip

\AtlasOrcid[0000-0002-6665-4934]{G.~Aad}$^\textrm{\scriptsize 102}$,
\AtlasOrcid[0000-0001-7616-1554]{E.~Aakvaag}$^\textrm{\scriptsize 17}$,
\AtlasOrcid[0000-0002-5888-2734]{B.~Abbott}$^\textrm{\scriptsize 122}$,
\AtlasOrcid[0000-0002-0287-5869]{S.~Abdelhameed}$^\textrm{\scriptsize 118a}$,
\AtlasOrcid[0000-0002-1002-1652]{K.~Abeling}$^\textrm{\scriptsize 54}$,
\AtlasOrcid[0000-0001-5763-2760]{N.J.~Abicht}$^\textrm{\scriptsize 48}$,
\AtlasOrcid[0000-0002-8496-9294]{S.H.~Abidi}$^\textrm{\scriptsize 30}$,
\AtlasOrcid[0009-0003-6578-220X]{M.~Aboelela}$^\textrm{\scriptsize 44}$,
\AtlasOrcid[0000-0002-9987-2292]{A.~Aboulhorma}$^\textrm{\scriptsize 36e}$,
\AtlasOrcid[0000-0001-5329-6640]{H.~Abramowicz}$^\textrm{\scriptsize 155}$
% ...
\bigskip
\\

% ...
$^{17}$Department for Physics and Technology, University of Bergen, Bergen; Norway.\\
% ...
$^{30}$Physics Department, Brookhaven National Laboratory, Upton NY; United States of America.\\
% ...
$^{36}$$^{(a)}$Facult\'e des Sciences Ain Chock, Universit\'e Hassan II de Casablanca;$^{(b)}$Facult\'{e} des Sciences, Universit\'{e} Ibn-Tofail, K\'{e}nitra;$^{(c)}$Facult\'e des Sciences Semlalia, Universit\'e Cadi Ayyad, LPHEA-Marrakech;$^{(d)}$LPMR, Facult\'e des Sciences, Universit\'e Mohamed Premier, Oujda;$^{(e)}$Facult\'e des sciences, Universit\'e Mohammed V, Rabat;$^{(f)}$Institute of Applied Physics, Mohammed VI Polytechnic University, Ben Guerir; Morocco.\\
% ...
$^{44}$Physics Department, Southern Methodist University, Dallas TX; United States of America.\\
% ...
$^{48}$Fakult\"{a}t Physik, Technische Universit{\"a}t Dortmund, Dortmund; Germany.\\
% ...
$^{54}$II. Physikalisches Institut, Georg-August-Universit\"{a}t G\"ottingen, G\"ottingen; Germany.\\
% ...
$^{102}$CPPM, Aix-Marseille Universit\'e, CNRS/IN2P3, Marseille; France.\\
% ...
$^{118}$$^{(a)}$New York University Abu Dhabi, Abu Dhabi;$^{(b)}$United Arab Emirates University, Al Ain; United Arab Emirates.\\
% ...
$^{122}$Homer L. Dodge Department of Physics and Astronomy, University of Oklahoma, Norman OK; United States of America.\\
% ...
$^{155}$Raymond and Beverly Sackler School of Physics and Astronomy, Tel Aviv University, Tel Aviv; Israel.\\
% ...
$^{a}$ Also at Affiliated with an institute formerly covered by a cooperation agreement with CERN.\\
$^{b}$ Also at An-Najah National University, Nablus; Palestine.\\
$^{c}$ Also at Borough of Manhattan Community College, City University of New York, New York NY; United States of America.\\
$^{d}$ Also at Center for Interdisciplinary Research and Innovation (CIRI-AUTH), Thessaloniki; Greece.\\
$^{e}$ Also at Centre of Physics of the Universities of Minho and Porto (CF-UM-UP); Portugal.\\
$^{f}$ Also at CERN, Geneva; Switzerland.\\
% ...

\end{flushleft}
